\documentclass{jfm}

\usepackage{graphicx}
\usepackage{natbib}
\usepackage{hyperref}
\usepackage{amsmath}
\usepackage{bm}
\usepackage{newtxtext,newtxmath}

\newcommand{\vect}[1]{\bm{#1}}

\newcommand{\aver}[1]{ \! \left\langle {#1} \right \rangle \!}
\newcommand{\paver}[1]{ \overline{ {#1} } }
\newcommand{\coh}[1]{ \tilde{ {#1} } }

\newcommand{\mf}[1]{{#1}^{mf}}
\newcommand{\ms}[1]{{#1}^{ms}}
\newcommand{\mc}[1]{{#1}^{mc}}
\newcommand{\cs}[1]{{#1}^{cs}}

\newcommand{\xit}{\Xi}  %t=averaged in time   %ph=phase by phase
\newcommand{\xiph}{\xi}
\newcommand{\phit}{\Phi}
\newcommand{\phiph}{\phi}
\newcommand{\psit}{\Psi}
\newcommand{\psiph}{\psi}

\newcommand{\pt}{P}
\newcommand{\pph}{p}
\newcommand{\dt}{D}
\newcommand{\dph}{d}
\newcommand{\pit}{\Pi}
\newcommand{\piph}{\pi}
\newcommand{\ft}{\mathrm{F}}
\newcommand{\fph}{\mathrm{f}}

\title[Structure functions with triple decomposition]
{Structure function tensor equations with triple decomposition}

\author[F.Gattere, A.Chiarini, E.Gallorini, M.Quadrio]
{Federica Gattere, Alessandro Chiarini, Emanuele Gallorini and Maurizio Quadrio}

\affiliation{Dipartimento di Scienze e Tecnologie Aerospaziali, Politecnico di Milano, via La Masa 34, 20156 Milano, Italy
}

\begin{document}
\maketitle

%%%%%%%%%%%%%%%%%%%%%%%%%%%%%%%%%%%%%%%%%%%%%%%%%%
\begin{abstract}
Exact budget equations are derived for the coherent and stochastic contributions to the second-order structure function tensor. They extend the anisotropic generalised Kolmogorov equations (AGKE) by considering the coherent and stochastic parts of the Reynolds stress tensor, and are useful for the statistical description of turbulent flows with periodic or quasi-periodic features, like e.g. the alternate shedding after a bluff body.
While the original AGKE describe production, transport, inter-component redistribution and dissipation of the Reynolds stresses in the combined space of scales and positions, the new equations, called $\varphi$AGKE, contain the phase $\varphi$ as an additional independent variable, and describe the interplay among the mean, coherent and stochastic fields at the various phases.
The newly derived $\varphi$AGKE are then applied to a case where an exactly periodic external forcing drives the flow: a turbulent plane channel flow modified by harmonic spanwise oscillations of the wall to reduce drag. 
The phase-by-phase action of the oscillating transversal Stokes layer generated by the forcing on the near-wall turbulent structures is observed, and a detailed description of the scale-space interaction among mean, coherent and stochastic fields is provided thanks to the $\varphi$AGKE.
\end{abstract}
%%%%%%%%%%%%%%%%%%%%%%%%%%%%%%%%%%%%%%%%%%%%%%%%%%

\section{Introduction}
\label{sec:intro}
Understanding the multiscale nature of turbulence and the sustaining mechanisms of turbulent fluctuations is a long-standing effort in fluid mechanics, motivated by the ambition to determine and possibly to manipulate the mean flow.
According to the classic arguments by Richardson and Kolmogorov, at large enough Reynolds numbers a clear scale separation is expected between the large energy-containing scales and the small dissipative ones. 
Fluctuations of different scales interact non-linearly, and a cascade mechanism transfers energy (on average) towards the dissipating scales. 
The geometrical information embedded in the larger scales vanishes at smaller ones, so that turbulence becomes locally isotropic below a small enough scale. 
However, in turbulent flows with practical interest, the scale separation is often incomplete, owing to the finite value of the Reynolds number and to the presence of boundaries; studying such flows is particularly challenging, because of their strongly anisotropic and inhomogeneous nature, which implies that the very concept of scale comes to depend on the position in the physical space.

Among the approaches developed over the years to describe anisotropic and inhomogeneous flows, the anisotropic generalised Kolmogorov equations, or AGKE, are well suited to account for the multiscale nature of turbulence. 
The AGKE \citep{gatti-etal-2020} are exact budget equations for each component of the second-order structure function tensor. They extend the generalised Kolmogorov equation or GKE \citep[see e.g.][]{hill-2001, danaila-etal-2001}, sometimes referred to as K\'arm\'an--Howarth--Monin--Hill equation \citep{alvesportela-papadakis-vassilicos-2017}, which, in turn, is the exact budget equation for half the trace of the second-order structure function tensor, i.e. the scale energy. 
The AGKE, which consider each tensor component separately, describe the production, inter-component redistribution, transport, and dissipation of the Reynolds stresses simultaneously across the scales and in the physical space.
Unlike the GKE, they fully account for anisotropy and inhomogeneity, and feature a pressure--strain term that plays a central role in redistribution. 
Moreover, the AGKE simplify the structural analysis of turbulence, owing to the direct link of each tensor component to the correlation function \citep{davidson-etal-2006, gatti-etal-2020}.

The GKE has been already applied to several flows to describe how inhomogeneity changes the Richardson--Kolmogorov scenario, possibly leading to inverse (from small to large scales) energy transfer: the plane channel flow at different $Re$ \citep{cimarelli-deangelis-casciola-2013, cimarelli-etal-2016}, the flow over a bump \citep{mollicone-etal-2018}, the wake of a square cylinder \citep{alvesportela-papadakis-vassilicos-2017} and the plane jet \cite{cimarelli-etal-2021}. 
Using GKE, \cite{yao-mollicone-papadakis-2022} showed that an intense inverse cascade dominates a boundary layer undergoing bypass transition.
\cite{danaila-voivenel-varea-2017} derived the variable-viscosity GKE and proved that, in flows with mixing of two or more fluids, all scales evolve in a similar fashion only for regions where viscosity is uniform.
\cite{lai-etal-2018} derived the variable-density GKE and studied the multi-material effects on the interscale energy transfers in a turbulent round jet, finding that the deformation of smaller turbulent eddies into larger ones accompanies energy transfers. 
\cite{arun-etal-2021} derived the budget equation for the derivative of the two-point velocity correlation for compressible flows, and identified the effects of variable density and dilatation on the energy cascades. The more recent AGKE, instead, have been first demonstrated in a plane channel flow \citep{gatti-etal-2020}, and then used to investigate the ascending/descending and direct/inverse cascades of the Reynolds stresses in a turbulent Couette flow \citep{chiarini-etal-2021} and to characterise the structure of turbulence in the flow past a rectangular cylinder \citep{chiarini-etal-2022}.

It is not uncommon to encounter turbulent flows in which large scales are relatively organised in space, and follow a temporally repeating pattern. 
This happens in presence of an external periodic forcing, or when the flow is quasi-periodic because of instabilities, as in the turbulent wake of bluff bodies. 
An example of the former class, which is considered in the second half of this paper as a simpler testbench, is the canonical turbulent channel flow modified by periodic spanwise wall oscillation to obtain skin-friction drag reduction \citep{jung-mangiavacchi-akhavan-1992}. 
The spanwise forcing creates a coherent periodic velocity field, known as the generalised Stokes layer \citep{quadrio-ricco-2011}, which superimposes on the stochastic turbulent fluctuations. 
The latter class includes the quasi-periodic K\'arm\'an-like vortices in the turbulent wake of bluff bodies, forming after the roll-up of the separating shear layers. 
Such quasi-periodic structures, usually referred to as coherent motions, interact with the stochastic fluctuations and affect their organisation. 

A complete, multiscale description of the interaction among the mean, the coherent (e.g. periodic) and the stochastic fields is highly desirable.
Indeed, one can resort to a triple decomposition of the velocity and pressure fields into mean, coherent and stochastic motions, and use it, together with the single-point Reynolds stress budget equations, to describe how these large-scale motions interact with the turbulent fluctuations in the physical space. 
For the spanwise-oscillating wall, \cite{agostini-touber-leschziner-2014} found that the phase variation of the stochastic contribution to the Reynolds stresses is mainly driven by production, and that the dissipation plays only a marginal role; they concluded that the increase of the dissipation can not be the cause of drag reduction. 
For the alternate shedding behind a bluff body, \cite{kiya-matsumura-1988} experimentally investigated the various frequency components of the stochastic motions in the wake behind a flat plate perpendicular to the flow. They found that the frequency of the main contributions to the stochastic shear stresses is one half of the vortex-shedding frequency, explaining it with the different spanwise arrangement of consecutive coherent vortices. 
In both cases, however, the description was incomplete: a triple decomposition alone does not capture the interaction between coherent and stochastic motions in the space of scales.

\cite{alvesportela-papadakis-vassilicos-2020} followed \cite{thiesset-danaila-antonia-2014} and used the GKE together with a triple decomposition to describe the interaction between the coherent and stochastic motions in the space of scales and positions. 
They arrived at two budget equations for the coherent and stochastic parts of the scale energy, and applied them to the turbulent wake past a square cylinder. 
Interestingly, they found that the mean flow does not feed the stochastic field directly, but it produces kinetic energy that feeds the large-scale coherent structures shed in the wake. 
Part of this energy is then transferred towards the stochastic turbulent fluctuations, at all scales. 
Although promising, the approach by \cite{alvesportela-papadakis-vassilicos-2020} is still affected by limitations, discussed by \cite{thiesset-danaila-2020}, that prevent a complete understanding of the  interaction among the three fields. 
This is because their budget equations are obtained by averaging over the phase of the coherent motions, and the phase dependence is lost in the process. 
Furthermore, being based on the GKE, their procedure considers only the scale energy, and does not describe the pressure--strain redistribution among the various components of the Reynolds stress tensor. 
Finally, \cite{alvesportela-papadakis-vassilicos-2020} additionally discard directional information by taking orientation averages of every term of the budget equations. 

The present work goes one step further to overcome these limitations. 
We use a triple decomposition to extend the AGKE, and arrive at two phase-by-phase budget equations for the coherent and stochastic parts of each component of the structure function tensor. 
These equations, named $\varphi$AGKE, describe the phase-by-phase mean-coherent-stochastic interaction of each component of the Reynolds stresses in the combined space of scales and positions. There is no phase-average involved, so that the description is complete.
The paper is structured as follows. After this introduction, in \S\ref{sec:math} we briefly recall the AGKE for the classic Reynolds decomposition and introduce the $\varphi$AGKE for the triple decomposition, discussing the meaning of the various terms.
In the second part of the contribution, in \S\ref{sec:tcf-ow}, we provide a relatively simple example, and apply the new budget equations to a turbulent channel flow subjected to an oscillatory spanwise wall motion, chosen because of the deterministic nature of the periodic component. In \S\ref{sec:interpretation} we demonstrate how the $\varphi$AGKE describe the mean-coherent-stochastic interaction, and shed light into the complex working mechanism of the oscillating wall. 
The paper closes with a brief discussion in \S\ref{sec:conclusions}. 
Appendix \ref{sec:eqderiv} contains the detailed derivation of the $\varphi$AGKE from the Navier--Stokes equations, followed in Appendix \ref{sec:tcf-phiagke} by their specialization to the plane channel flow with oscillating walls. In Appendix \ref{sec:ens-aver} the velocity field induced by the ensemble-averaged quasi-streamwise vortex at different phases is computed and used to support the $\varphi$AGKE-based analysis of the channel flow with oscillating walls.

\section{Mathematical formulation}
\label{sec:math}
In this Section we introduce the triple decomposition and recall briefly the standard AGKE, before presenting the new $\varphi$AGKE, whose detailed derivation is reported in Appendix \ref{sec:eqderiv}.

%------------------------------------------------------ 
\subsection{Triple decomposition of the velocity field}

An incompressible turbulent flow, varying in space $\vect{x}$ and time $t$, is typically described via its mean and fluctuating velocity and pressure fields, defined after the classic Reynolds decomposition. Provided the flow exhibits well-defined non-stochastic (e.g. periodic) features, the fluctuating field can be further decomposed into a coherent and a stochastic part. Therefore, the velocity field reads:
\begin{equation}
\vect{u} = \vect{U} + \underbrace{\coh{\vect{u}} + \vect{u}''}_{\vect{u}'},
\label{eq:triple_decomposition}
\end{equation}
where $\vect{U}$, $\vect{u}'$, $\coh{\vect{u}}$ and $\vect{u}''$ indicate the mean, fluctuating, coherent and stochastic parts of the velocity field $\vect{u}$. The mean velocity $\vect{U}$ is defined as $\vect{U} \equiv \aver{\vect{u}}$, with the operator $\aver{\cdot}$ indicating ensemble averaging, which under the ergodic hypothesis becomes equivalent to averaging over homogeneous directions and time (if the flow is statistically stationary). 
For a single realisation without homogeneous directions, the mean is simply a temporal average:
\begin{equation}
\vect{U}(\vect{x}) \equiv \lim_{\tau \to +\infty} \frac{1}{\tau} \int_0^\tau \vect{u}(\vect{x},t) dt . 
\end{equation}

Considering a periodic motion with period $T$ and phase $\varphi \in (0,2\pi]$, the overbar $\paver{\cdot}$ denotes the phase average operator over an integer number $N$ of periods. Like $\aver{\cdot}$, it includes averaging over the homogeneous directions. Considering again a single realisation without homogeneous directions, $\paver{\cdot}$ is defined as:
\begin{equation}
\paver{\vect{u}}(\vect{x},\varphi) \equiv \lim_{N\to +\infty} \frac{1}{N}  \sum_{n = 0}^{N-1} \vect{u}\left(\vect{x}, \left( \frac{\varphi}{2 \pi} + n \right) T \right).
\end{equation}

The coherent field $\coh{\vect{u}}$ is thus defined as 
\[
\coh{\vect{u}}(\vect{x},\varphi) = \paver{\vect{u}}(\vect{x},\varphi) - \vect{U}(\vect{x}) ,
\]
and the stochastic vector field $\vect{u}''$ is defined after the triple decomposition \eqref{eq:triple_decomposition} as $\vect{u}''=\vect{u}-\vect{U}-\coh{\vect{u}}$. 
An analogous triple decomposition is used to decompose the pressure field $p=P+\coh{p}+p''$, with $\coh{p}+p''=p'$.

%-------------------------------------------------------------------
\subsection{The anisotropic generalised Kolmogorov equations (AGKE)}
\label{sec:AGKE}

Before presenting the $\varphi$AGKE, the standard AGKE based on the Reynolds' decomposition are recalled. Full details on their derivation from the incompressible Navier--Stokes equations are provided by \cite{gatti-etal-2020}.

\begin{figure}
\centering
\includegraphics[width=0.8\textwidth]{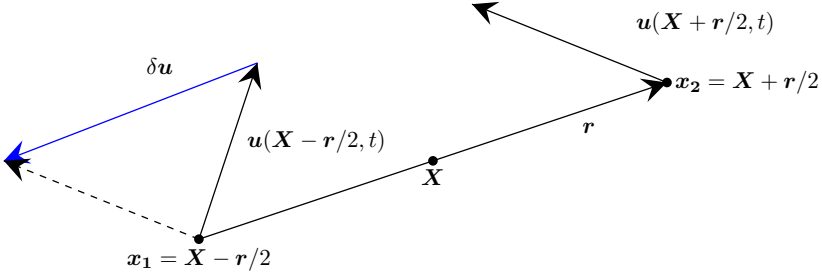}
\caption{Sketch of two points $\vect{x}_1$ and $\vect{x}_2$ involved in the definition of the second-order structure function tensor. $\vect{X}=( \vect{x}_1 + \vect{x}_2 )/2$ and $\vect{r}=\vect{x}_2-\vect{x}_1$ indicate their mid-point and separation vector, respectively. $\delta \vect{u} = \vect{u}_2 - \vect{u}_1$ is the velocity increment between the two points.}
\label{fig:sfsketch}
\end{figure}

Exact budget equations can be written for the components of the second-order structure function tensor $\aver{\delta u_i \delta u_j}$, where $\delta u_i=u_i(\vect{X}+\vect{r}/2,t)-u_i(\vect{X}-\vect{r}/2,t)$ is the $i-$th component of the velocity difference between two points $\vect{x}_1$ and $\vect{x}_2$, identified by their midpoint $\vect{X} = (\vect{x}_1+\vect{x}_2)/2$ and their separation vector $\vect{r} = (\vect{x}_2-\vect{x}_1)$, as shown by the sketch in figure \ref{fig:sfsketch}. The Reynolds' decomposition leads to budget equations for $\delta U_i \delta U_j$ and $\aver{\delta u'_i \delta u'_j}$. 
In general, the time-independent tensor $\delta U_i \delta U_j$ depends upon six independent variables, i.e. the six coordinates of $\vect{X}$ and $\vect{r}$. The tensor $\aver{\delta u'_i \delta u'_j}$ additionally features time $t$ as an independent variable if the process is not statistically stationary (e.g. periodic), and is related to the Reynolds stresses $\aver{u'_i u'_j}$ and to the spatial correlation tensor $R_{ij}$ \citep{davidson-etal-2006,agostini-leschziner-2017} as
\begin{equation}
\aver{\delta u'_i \delta u'_j}(\vect{X},\vect{r},t) =  V_{ij}(\vect{X},\vect{r},t) - R_{ij}(\vect{X},\vect{r},t) - R_{ij}(\vect{X},-\vect{r},t)
\end{equation}
where 
\begin{equation}
V_{ij}(\vect{X},\vect{r},t)=\aver{u'_i u'_j} \left( \vect{X} + \frac{\vect{r}}{2}, t \right) + \aver{u'_i u'_j} \left( \vect{X} - \frac{\vect{r}}{2}, t \right)
\end{equation}
is the sum of the single-point Reynolds stresses evaluated at the two points $\vect{X} \pm \vect{r}/2$, and
\begin{equation}
R_{ij}(\vect{X},\vect{r},t) =  \aver{ u'_i \left(\vect{X}+\frac{\vect{r}}{2},t \right) u'_j \left(\vect{X}-\frac{\vect{r}}{2},t \right)}
\label{eq:corr}
\end{equation}
is the two-points spatial correlation function. 

The budget equations for the components of the mean second-order structure function tensor $\delta U_i \delta U_j$ are presented here for the first time; they were not reported by \cite{gatti-etal-2020}, and the tensor has received little attention so far, owing to its irrelevance in homogeneous isotropic turbulence, where there is no mean flow. The mean AGKE are written compactly as
\begin{equation}
\frac{\partial \phit^m_{k,ij}}{\partial r_k} +
\frac{\partial \psit^m_{k,ij}}{\partial X_k} = 
\xit^m_{ij} ,
\label{eq:AGKEmean}
\end{equation}
where the repeated index $k$ implies summation. The following notation is adopted. Uppercase letters (e.g. $\phit$, $\psit$ and $\xit$) will be used to denote time-averaged quantities, and lowercase letters (e.g. $\phiph$, $\psiph$ and $\xiph$) for phase-dependent quantities. Furthermore, superscripts $m$, $f$, $c$ and $s$ are used to label terms in the budget equations for the mean structure function tensor $\delta U_i \delta U_j$, the fluctuating structure function tensor $\aver{\delta u'_i \delta u'_j}$, the coherent structure function tensor $\paver{\delta \coh{u}_i \delta \coh{u}_j} = \delta \coh{u}_i \delta \coh{u}_j$, and the stochastic structure function tensor $\paver{\delta u''_i \delta u''_j}$.

The fluxes $\Phi^m_{k,ij}$ and $\Psi^m_{k,ij}$ are the mean scale- and physical-space fluxes, i.e.
\begin{equation}
\begin{split}
\phit^m_{k,ij}      &=\underbrace{ \delta U_k \delta U_i \delta U_j }_{ \text{Mean transport} }   
               + \underbrace{ \delta U_j \aver{ \delta u'_k \delta u'_i } + \delta U_i \aver{ \delta u'_k \delta u'_j }  }_{ \text{Fluctuating transport} } %\nonumber\\ 
                    \underbrace{ - 2 \nu \frac{ \partial  \delta U_i \delta U_j}{ \partial r_k} }_{ \text{Viscous diffusion} } \ \ \ k = 1,2,3
\end{split}
\label{eq:mean-phi}           
\end{equation}
and
\begin{equation}
\begin{split}
\psit^m_{k,ij}      &=\underbrace{  U_k^* \delta U_i \delta U_j }_{ \text{Mean transport} }   
               + \underbrace{ \delta U_j \aver{ u'^*_k \delta u'_i } + \delta U_i \aver{  u'^*_k \delta u'_j }  }_{ \text{Fluctuating transport} }
                \underbrace{+ \frac{1}{\rho} \delta P \delta U_j  \delta_{ki}  + \frac{1}{\rho} \delta P \delta U_i  \delta_{kj} }_{ \text{Pressure transport} }  + \\
                & \underbrace{ -  \frac{\nu}{2} \frac{ \partial  \delta U_i \delta U_j}{ \partial X_k} }_{ \text{Viscous diffusion} } \ \ \ k = 1,2,3.
\end{split}
\label{eq:mean-psi}           
\end{equation}
where $\delta_{ij}$ is the Kronecker delta, $\nu$ is the kinematic viscosity, and the asterisk superscript $(\cdot)^{\ast}$ indicates the arithmetic average of a quantity between the two points $\vect{X} \pm \vect{r}/2$. 
The term $\xit^m_{ij}$, instead, is the mean source and reads
\begin{equation}
\label{eq:mean-xi}
\begin{split}
\xit^m_{ij}   & =
                 - \underbrace{\Bigl[- \aver{u^{\prime  *}_k \delta u'_j } \delta \left( \frac{ \partial U_i }{ \partial x_k } \right)
                 -\aver{u^{\prime  *}_k \delta u'_i} \delta \left( \frac{ \partial U_j }{ \partial x_k } \right)
                 -\aver{\delta u'_k \delta u'_j } \left( \frac{\partial U_i}{\partial x_k} \right)^*- \aver{\delta u'_k \delta u'_i } \left( \frac{\partial U_j}{\partial x_k} \right)^* \Bigr]}_{\text{Mean-fluctuating production} \  (\mf{\pt}_{ij})}+ \\                    
                 &  \underbrace{ + \frac{1}{\rho}  \delta P \frac{\partial \delta U_i}{\partial X_j} + \frac{1}{\rho}  \delta P \frac{\partial \delta U_j}{\partial X_i} }_{\text{Pressure strain}  \ (\pit^m_{ij})} 
                    \underbrace{- 4 \epsilon_{ij}^{m*} }_{\text{Dissipation} \ (\dt^m_{ij})} + \underbrace{\delta U_j \delta F_i + \delta U_i \delta F_j.}_{\text{Forcing interaction} \ (\ft^m_{ij})}\\
\end{split}  
\end{equation}

The standard AGKE, presented by \cite{gatti-etal-2020}, pertain to increments of the fluctuating velocity field, and describe the production, transport, redistribution and dissipation of each component, in the physical space $\vect{X}$ and in the space of scales $\vect{r}$. They can be written compactly as:
\begin{equation}
  \frac{\partial \aver{\delta u'_i \delta u'_j} }{\partial t} +
  \frac{\partial \phit^f_{k,ij}}{\partial r_k} +
  \frac{\partial \psit^f_{k,ij}}{\partial X_k} = 
  \xit^f_{ij} .
\label{eq:agke}
\end{equation}

The scale-space fluxes $\phit^f_{k,ij}$ and physical-space fluxes $\psit^f_{k,ij}$ are defined as:
\begin{equation}
\phit^f_{k,ij} =\underbrace{\aver{\delta U_k \delta u'_i \delta u'_j } }_{\text{Mean transport}} +  
              \underbrace{\aver{\delta u'_k \delta u'_i \delta u'_j} }_{\text{Fluctuating transport}}  
              \underbrace{ -2 \nu \frac{\partial}{\partial r_k} \aver{\delta u'_i \delta u'_j}}_{\text{Viscous diffusion}} \ \ \ k=1,2,3
\label{eq:phi_r_gen}
\end{equation}
and
\begin{equation}
\begin{split}
\psit^f_{k,ij} & =  \underbrace{ \aver{ U_k^{\ast} \delta u'_i \delta u'_j} }_{\text{Mean transport}}+
                \underbrace{ \aver{u_k'^{\ast} \delta u'_i \delta u'_j} }_{\text{Fluctuating transport}} + 
                \underbrace{\frac{1}{\rho} \aver{\delta p' \delta u'_i} \delta_{kj} + \frac{1}{\rho} \aver{\delta p' \delta u'_j} \delta_{ki}}_{\text{Pressure transport}}+\\
                &\underbrace{- \frac{\nu}{2} \frac{\partial}{\partial X_k} \aver{\delta u'_i \delta u'_j}}_{\text{Viscous diffusion}}  \ \ \ k=1,2,3.
\label{eq:phi_c_gen}
\end{split}
\end{equation}

The term $\xit^f_{ij}$ in \eqref{eq:agke} is the source for $\aver{ \delta u'_i \delta u'_j}$ and reads:
\begin{equation}
\begin{split}
\xit^f_{ij} =& \underbrace{-\aver{u_k'^{\ast} \delta u'_j} \delta \left( \frac{ \partial U_i}{\partial x_k} \right) - \aver{u_k'^{\ast} \delta u'_i} \delta \left(\frac{\partial U_j}{ \partial x_k}\right)
-\aver{\delta u'_k \delta u'_j} \left(\frac{ \partial U_i}{ \partial x_k}\right)^{\ast} - \aver{\delta u'_k \delta u'_i} \left( \frac{\partial U_j}{ \partial x_k} \right)^{\ast}}_{\text{Mean-fluctuating production} (\mf{\pt}_{ij})} +\\
          & + \underbrace{\frac{1}{\rho}\aver{\delta p' \frac{ \partial \delta u'_i}{ \partial X_j}} + \frac{1}{\rho}\aver{\delta p' \frac{\partial \delta u'_j}{ \partial X_i}}}_{\text{Pressure strain } (\pit^f_{ij})} %
            \underbrace{-4 \epsilon_{ij}^{f \ast} }_{\text{Dissipation } (\dt^f_{ij})} + \underbrace{\aver{\delta u'_j \delta f'_i} + \aver{\delta u'_i \delta f'_j}}_{\text{Forcing interaction} (\ft^f_{ij})} \,.
\end{split}
\label{eq:xi_gen}
\end{equation}
in which $\epsilon_{ij}^f$ is the pseudo-dissipation tensor $\aver{\partial u_i'/\partial x_k \partial u_j' / \partial x_k}$. 
The source term $\xit^f_{ij}$ identifies scales and positions with a net sink ($\xit^f_{ij}<0$) or a net source ($\xit^f_{ij}>0$) for each component of the Reynolds stresses. 
The separation of $\xit^f_{ij}$ in its constituent terms provides insight on mean-fluctuating production $\mf{\pt}_{ij}$ (which also appears in \eqref{eq:mean-xi} with opposite sign), redistribution $\pit^f_{ij}$, dissipation $\dt^f_{ij}$ and interaction with external fluctuating volume forces $\ft^f_{ij}$ of turbulent stresses among scales and positions (note that the forcing interaction term was missing in the original AGKE formulated by \cite{gatti-etal-2020}). The flux vectors describe the various transfer processes, and their field lines visualise how fluctuations are transferred among scales and positions, via direct and inverse cascades. 
It should be recalled that, as stressed by \cite{gatti-etal-2020}, when interpreting AGKE results to extract structural turbulence information, local peaks of the structure functions always need to be connected to local maxima/minima of the correlation functions whenever a separation along an inhomogeneous direction is involved.

%------------------------------------------------
\subsection{The phase-aware AGKE, or $\varphi$AGKE}

By using the triple decomposition \eqref{eq:triple_decomposition}, the phase-averaged fluctuating structure function tensor $ \paver{\delta u'_i \delta u'_j}(\vect{X},\vect{r},\varphi) $ can be separated into its coherent and stochastic parts, i.e.
\begin{equation}
\paver{\delta u_i' \delta u_j'}(\vect{X},\vect{r},\varphi) = 
\paver{\delta \coh{u}_i \delta \coh{u}_j} (\vect{X},\vect{r},\varphi) + \paver{\delta u''_i \delta u''_j} (\vect{X},\vect{r},\varphi);
\end{equation}
note that $\paver{\delta \coh{u}_i \delta \coh{u_j}} \equiv \delta \coh{u}_i \delta \coh{u_j}$ owing to the definition of the phase-average operator.
Two budget equations, called $\varphi$AGKE, can be written for $\delta \coh{u}_i \delta \coh{u}_j$ and $\paver{\delta u''_i \delta u''_j}$, which include, unlike the standard AGKE, the interplay among the mean, coherent and stochastic fields at each phase $\varphi$. 
These new equations extend in a significant way the work made by \cite{thiesset-danaila-antonia-2014} and \cite{alvesportela-papadakis-vassilicos-2020}, that considered the budget equations for $\aver{\delta \coh{u}_i \delta \coh{u}_i}(\vect{X},\vect{r})$ and $\aver{\delta u''_i \delta u''_i}(\vect{X},\vect{r})$.
They applied the triple decomposition to the trace $\aver{\delta u'_i \delta u'_i}$ of the second-order structure function tensor, instead of considering the whole tensor. The major difference, though, is that the dependence on the phase $\varphi$ of the coherent motion (or external forcing) was lost, because of the use of the $\aver{\cdot}$ operator. On the contrary, the $\varphi$AGKE retain full phase information.

The step-by-step derivation of the $\varphi$AGKE from the incompressible Navier--Stokes equations is described in Appendix \ref{sec:eqderiv}. 
At each phase $\varphi$, they link the phase variation of each component of the coherent and stochastic structure function tensors, at a given scale $\vect{r}$ and position $\vect{X}$, to the unbalance among inter-component redistribution, scale-space transport, dissipation and mean-coherent-stochastic interaction. The last term is obviously absent in the classic AGKE. 

The equations for the coherent and stochastic parts can be compactly written as:
\begin{equation}
\frac{2\pi}{T} \frac{\partial \delta \coh{u}_i \delta \coh{u}_j }{ \partial \varphi} +
\frac{\partial \phiph^c_{k,ij}}{\partial r_k} +
\frac{\partial \psiph^c_{k,ij}}{\partial X_k} = 
\xiph^c_{ij} + \zeta^c_{ij}
\label{eq:AGKEcoh}
\end{equation}
and 
\begin{equation}
\frac{2\pi}{T} \frac{\partial \paver{\delta u''_i \delta u''_j}}{\partial \varphi} +
\frac{\partial \phiph^s_{k,ij}}{\partial r_k} +
\frac{\partial \psiph^s_{k,ij}}{\partial X_k} =
\xiph^s_{ij} ,
\label{eq:AGKEsto}
\end{equation}
where, as above, the repeated index $k$ implies summation.

The first term in equations \eqref{eq:AGKEcoh} and \eqref{eq:AGKEsto} represents the phase variation of the coherent and stochastic components of the structure function tensor.
The coherent and stochastic scale fluxes $\phiph^c_{k,ij}$ and $\phiph^s_{k,ij}$, i.e. the fluxes of $\delta \coh{u}_i \delta \coh{u}_j$ and $\paver{\delta u''_i \delta u''_j}$ in the space of scales, are defined as:
\begin{equation}
\phiph^c_{k,ij}=\underbrace{ \delta U_k \delta \coh{u}_i \delta \coh{u}_j }_{\text{Mean transport}} + 
                    \underbrace{ \delta \coh{u}_k \delta \coh{u}_i \delta \coh{u}_j }_{ \text{Coherent transport} } +                   \underbrace{\paver{ \delta u''_k \delta u''_i } \delta \coh{u}_j 
                    + \paver{ \delta u''_k \delta u''_j } \delta \coh{u}_i }_{ \text{Stochastic transport} } 
                   \underbrace{-2 \nu \frac{ \partial  \delta \coh{u}_i \delta \coh{u}_j }{\partial r_k} }_{ \text{Viscous diffusion} }  \ \ \ k = 1,2,3  \label{eq:coh-phi} 
\end{equation}
and 
\begin{equation}
\phiph^s_{k,ij}      =\underbrace{ \delta U_k \paver{ \delta u''_i \delta u''_j} }_{ \text{Mean transport} } 
               + \underbrace{ \delta \coh{u}_k \paver{ \delta u''_i \delta u''_j }  }_{ \text{Coherent transport} }   
               + \underbrace{ \paver{\delta u''_k \delta u''_i \delta u''_j} }_{ \text{Stochastic transport} } 
                    \underbrace{ - 2 \nu \frac{ \partial \paver{ \delta u''_i \delta u''_j}}{ \partial r_k} }_{ \text{Viscous diffusion} } \ \ \ k = 1,2,3.
\label{eq:phi}           
\end{equation}
The coherent and stochastic spatial flux terms $\psiph^c_{k,ij}$ and $\psiph^s_{k,ij}$, i.e. the fluxes of $\delta \coh{u}_i \delta \coh{u}_j$ and $\paver{\delta u''_i \delta u''_j}$ in the physical space, are defined as:
\begin{align} 
\psiph^c_{k,ij} &=\underbrace{  U_k^* \delta \coh{u}_i \delta \coh{u}_j }_{ \text{Mean transport} } + 
              \underbrace{  \coh{u}_k^* \delta \coh{u}_i \delta \coh{u}_j  }_{ \text{Coherent transport} } +
                   \underbrace{  \paver{ u^{\prime  \prime * }_k \delta u^{\prime \prime}_i } \delta \coh{u}_j  +  \paver{ u^{\prime \prime *}_k \delta u''_j } \delta \coh{u}_i  }_{ \text{Stochastic transport} } + \nonumber
                 \underbrace{ \frac{1}{\rho} \delta \coh{p} \delta \coh{u}_i  \delta_{kj} }_{ \text{Pressure transport} } + \\
     &          \underbrace{+ \frac{1}{\rho} \delta \coh{p} \delta \coh{u}_j  \delta_{ki} }_{ \text{Pressure transport} } 
              \underbrace{ - \frac{\nu}{2} \frac{\partial  \delta \coh{u}_i \delta \coh{u}_j }{ \partial X_k } }_{ \text{Viscous diffusion} }  
      \ \ \ k = 1,2,3 \label{eq:coh-psi} 
\\ 
\psiph^s_{k,ij}       &=\underbrace{  U_k^* \paver{ \delta u''_i \delta u''_j } }_{ \text{Mean transport} } 
                  + \underbrace{  \coh{u}_k^* \paver{\delta u''_i \delta u''_j} }_{ \text{Coherent transport} }
                  + \underbrace{ \paver{ u_k^{\prime  \prime * } \delta u''_i \delta u''_j } }_{ \text{Stochastic transport} } + \nonumber            
                  \underbrace{ \frac{1}{\rho}\paver{ \delta p'' \delta u''_i} \delta_{kj}   +
                     \frac{1}{\rho}\paver{ \delta p'' \delta u''_j} \delta_{ki} }_{ \text{Pressure transport} } + \\
                &    \underbrace{ - \frac{\nu}{2} \frac{\partial \paver{\delta u''_i \delta u''_j}}{\partial X_k } }_{ \text{Viscous diffusion} }\ \ \ k = 1,2,3.            \label{eq:psi} 
\end{align}

The differences with the fluxes \eqref{eq:phi_r_gen} and \eqref{eq:phi_c_gen} appearing in the standard AGKE are worth noticing. Two new terms appear here to account for the effect of the coherent field upon transport in the stochastic field, labelled as coherent transport in equations \eqref{eq:phi} and \eqref{eq:psi}. Vice versa, how the stochastic field affects transport in the coherent field is reflected by the stochastic transport term in equations \eqref{eq:coh-phi} and \eqref{eq:coh-psi}). 

The coherent and stochastic source terms $\xiph^c_{ij}$ and $\xiph^s_{ij}$ denote the scale-space net production of $\delta \coh{u}_i \delta \coh{u}_j$ and $\paver{\delta u''_i \delta u''_j}$. They can be either positive or negative, and read:
\begin{equation}
\label{eq:coh-xi}
\begin{aligned}
\xiph^c_{ij}   & =\underbrace{- \delta \coh{u}_j \delta \coh{u}_k \left( \frac{\partial U_i}{\partial x_k} \right)^* 
                                - \delta \coh{u}_i \delta \coh{u}_k \left( \frac{\partial U_j}{\partial x_k} \right)^* 
                 -        \delta \coh{u}_j \coh{u}_k^* \delta \left( \frac{ \partial U_i }{ \partial x_k } \right) 
                 -        \delta \coh{u}_i \coh{u}_k^* \delta \left( \frac{ \partial U_j }{ \partial x_k } \right)}_{\text{Mean-coherent production} \  (\mc{\pph}_{ij})}+ \\
    & - \underbrace{\left[- \paver{ \delta u''_j \delta u''_k} \left( \frac{\partial \coh{u}_i}{\partial x_k} \right)^*
                 -  \paver{ \delta u''_i \delta u''_k} \left( \frac{\partial \coh{u}_j}{\partial x_k} \right)^*
                 -  \paver{ \delta u''_j u^{\prime \prime *}_k } \delta \left( \frac{ \partial \coh{u}_i }{\partial x_k}  \right) 
                 -  \paver{ \delta u''_i u^{\prime \prime *}_k } \delta \left(  \frac{ \partial \coh{u}_j }{\partial x_k} \right)   \right] }_{\text{Coherent-stochastic production} \ (\cs{\pph}_{ij})} +  \\                     
                 &  \underbrace{+ \frac{1}{\rho}  \delta \coh{p} \frac{\partial \delta \coh{u}_i}{\partial X_j} + \frac{1}        {\rho}  \delta \coh{p} \frac{\partial \delta \coh{u}_j}{\partial X_i} }_{\text{Pressure strain}  \ (\piph^c_{ij})} 
                    \underbrace{- 4 \epsilon_{ij}^{c \ast} }_{\text{Dissipation} \ (\dph^c_{ij})} + \underbrace{\delta \coh{u}_j \delta \coh{f}_i + \delta \coh{u}_i \delta \coh{f}_j}_{\text{Forcing interaction} (\fph^c_{ij})}\\
\end{aligned}  
\end{equation}
\begin{equation}
\begin{aligned}                 
\label{eq:xi}                    
\xiph^s_{ij} & =\underbrace{- \paver{ \delta u''_j \delta u''_k } \left( \frac{\partial U_i}{\partial x_k} \right)^* 
                 - \paver{ \delta u''_i \delta u''_k } \left( \frac{\partial U_j}{\partial x_k} \right)^* 
                 - \paver{ \delta u^{\prime \prime }_j u^{''* }_k} \delta \left( \frac{ \partial U_i }{ \partial x_k } \right) 
                 - \paver{ \delta u''_i u^{\prime \prime *}_k } \delta \left( \frac{ \partial U_j }{ \partial x_k } \right)}_{\text{Mean-stochastic production} \ (\ms{\pph}_{ij})} +\\
                 &  +\underbrace{\left[ - \paver{ \delta u''_j \delta u''_k} \left( \frac{\partial \coh{u}_i}{\partial x_k}  \right)^*
                 - \paver{ \delta u''_i \delta u''_k} \left(  \frac{\partial \coh{u}_j}{\partial x_k} \right)^*
                 - \paver{ \delta u''_j u^{\prime \prime *}_k } \delta \left(  \frac{ \partial \coh{u}_i }{\partial x_k} \right)
                 - \paver{ \delta u''_i u^{\prime \prime *}_k } \delta \left(  \frac{ \partial \coh{u}_j }{\partial x_k} \right) \right] }_{\text{Coherent-stochastic production} \ (\cs{\pph}_{ij})}  +\\                     
                 &  \underbrace{+ \frac{1}{\rho} \paver{ \delta p'' \frac{\partial \delta u''_i}{\partial X_j}} + \frac{1}{\rho} \paver{ \delta p'' \frac{\partial \delta u''_j}{\partial X_i} }}_{\text{Pressure strain} \ (\piph^s_{ij})} 
                    \underbrace{- 4 \epsilon_{ij}^{s \ast}}_{\text{Dissipation} \ (\dph^s_{ij})} + \underbrace{\paver{\delta u''_j \delta f''_i} + \paver{\delta u''_i \delta f''_j}}_{\text{Forcing interaction} (\fph^s_{ij})}.                   
\end{aligned}
\end{equation}

Among the terms appearing in the source, the mean-coherent and mean-stochastic productions $\mc{\pph}_{ij}$ and $\ms{\pph}_{ij}$ indicate the scales and positions where the mean flow feeds, or drains energy from, the coherent and stochastic fields: they are not positive definite, and therefore can be either sources or sinks.
They both contribute to the mean-fluctuating production $\mf{\pt}_{ij}$ in equation \eqref{eq:mean-xi}, as $\mf{\pt}_{ij}=\aver{\mc{\pph}_{ij}}+\aver{\ms{\pph}_{ij}}$.  
The coherent-stochastic production $\cs{\pph}_{ij}$ indicates the exchange of stresses between the coherent and stochastic fields, and appears in the budgets for $\delta \coh{u}_i \delta \coh{u}_j$ and $\paver{\delta u''_i \delta u''_j}$ with opposite sign. 
$\dph^c_{ij}$ and $\dph^s_{ij}$ denote viscous dissipation, and the pressure-strain terms $\piph^c_{ij}$ and $\piph^s_{ij }$ describe the interplay between pressure and velocity fields. 
Pressure--strain terms involve neither production nor dissipation of energy, and no cross-talk between coherent and fluctuating fields. 
Overall, among the source terms, the productions $\mc{\pph}_{ij}$, $\ms{\pph}_{ij}$ and $\cs{\pph}_{ij}$ are the only ones that connect the mean, coherent and fluctuating budgets, and are essential to ascertain how the mean, stochastic and coherent fields force each other. The forcing interactions $\fph^c_{ij}$ and $\fph^s_{ij}$ represent the power injected into the system by the interaction of a coherent and stochastic external volume forcing with the coherent and stochastic flow fields, respectively.

Finally, in equation \eqref{eq:AGKEcoh} for $\delta \coh{u}_i \delta \coh{u}_j$ a new term $\zeta^c_{ij}$ appears on the right-hand side. It describes the inter-phase interaction driven by the coherent flow field, and is defined as:
\begin{equation}
\begin{aligned}
\zeta^c_{ij} & =  \frac{\partial }{\partial r_k} 
                      \left[ \aver{\delta \coh{u}_i \delta \coh{u}_k} \delta \coh{u}_j + \aver{\delta \coh{u}_j \delta \coh{u}_k} \delta \coh{u}_i \right]  
                    + \frac{\partial}{\partial X_k} \left[ \aver{\coh{u}_k^* \delta \coh{u}_i } \delta \coh{u}_j + \aver{\coh{u}_k^* \delta \coh{u}_j } \delta \coh{u}_i \right] + \\
                    &+ \frac{\partial }{\partial r_k} 
                      \left[ \aver{\delta u''_i \delta u''_k} \delta \coh{u}_j + \aver{\delta u''_j \delta u''_k} \delta \coh{u}_i \right]  
                    + \frac{\partial}{\partial X_k} \left[ \aver{u''^*_k \delta u''_i } \delta \coh{u}_j + \aver{u''^*_k \delta u''_j } \delta \coh{u}_i \right] + \\
                 & - \aver{\delta \coh{u}_i \delta \coh{u}_k } \left( \frac{\partial \coh{u}_j}{\partial x_k} \right)^* -
                  \aver{\delta \coh{u}_j \delta \coh{u}_k } \left( \frac{\partial \coh{u}_i}{\partial x_k} \right)^* -
                  \aver{\delta \coh{u}_i \coh{u}_k^* } \delta \left( \frac{\partial \coh{u}_j}{\partial x_k} \right) -
                  \aver{\delta \coh{u}_j \coh{u}_k^* } \delta \left( \frac{\partial \coh{u}_i}{\partial x_k} \right)+\\
                   & - \aver{\delta u''_i \delta u''_k } \left( \frac{\partial \coh{u}_j}{\partial x_k} \right)^* -
                  \aver{\delta u''_j \delta u''_k } \left( \frac{\partial \coh{u}_i}{\partial x_k} \right)^* -
                  \aver{\delta u''_i u''^*_k } \delta \left( \frac{\partial \coh{u}_j}{\partial x_k} \right) -
                  \aver{\delta u''_j u''^*_k } \delta \left( \frac{\partial \coh{u}_i}{\partial x_k} \right).
\end{aligned}
\end{equation}

The terms in the last two rows above resemble a production term, and indicate the production of $\delta \coh{u}_i \delta \coh{u}_j$ due to the correlation of each phase with all the others. 

By averaging equations \eqref{eq:AGKEcoh} and \eqref{eq:AGKEsto} over the phases, the budget equations for $\aver{\delta \coh{u}_i \delta \coh{u}_j}(\vect{X},\vect{r})$ and $\aver{\delta u''_i \delta u''_j}(\vect{X},\vect{r})$ are obtained. 
In doing this, the inter-phase contributions vanish, since by definition they have zero average. 
The sum of the equations for the three diagonal components of $\aver{\delta \coh{u}_i \delta \coh{u}_j}$ and $\aver{\delta u''_i \delta u''_j}$ yields  the GKE equations used by \cite{alvesportela-papadakis-vassilicos-2020}. 
If the equations for $\aver{\delta \coh{u}_i \delta \coh{u}_j}$ and $\aver{\delta u''_i \delta u''_j}$ are added together, the standard AGKE for the fluctuating field $\aver{\delta u'_i \delta u'_j}$ are recovered. 

\section{Turbulent drag reduction by the spanwise-oscillating wall}
\label{sec:tcf-ow}

The $\varphi$AGKE are now applied to a fully developed turbulent channel flow subjected to a spanwise harmonic oscillation of the walls. 
This flow is a convenient example where the deterministic external periodic forcing provides an unambiguous definition of the phase, yet the physics behind drag reduction is interesting and not fully understood yet. 

\begin{figure}
\centering
\includegraphics[width=1\textwidth]{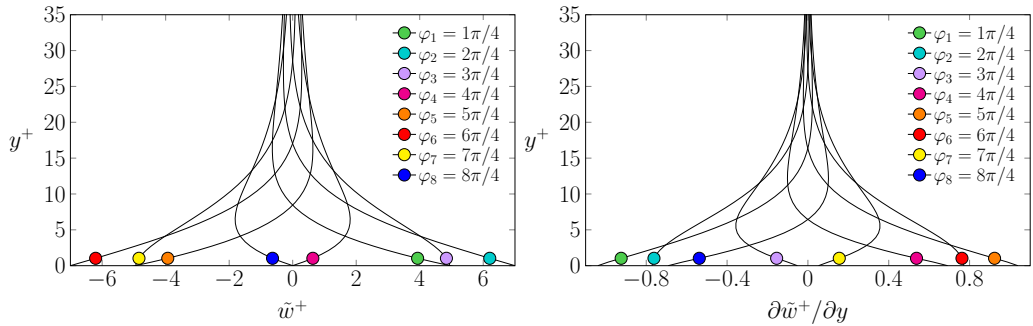}
\caption{Wall-normal profile of the spanwise coherent velocity $\coh{w}^+$ (left) and shear $\partial \coh{w}^+/\partial y$ (right), plotted at 8 equally spaced phases $\varphi_1, \ldots \varphi_8$ along the period $T^+=250$.}
\label{fig:channel}
\end{figure}

The spanwise oscillating wall is a well-known skin-friction drag reduction technique, intensely studied over the last thirty years \citep[see][and references therein]{ricco-skote-leschziner-2021}. The channel walls periodically move along the spanwise direction, according to:
\begin{equation}
w_w(t) = A \sin \left( \frac{2 \pi}{T} t \right),
\label{eq:OW}
\end{equation}
where $A$ and $T$ are the prescribed amplitude and period of the sinusoidal oscillation, and $w_w$ is the spanwise velocity of the wall. $x$, $y$ and $z$ ($u$, $v$ and $w$) denote the streamwise, wall-normal and spanwise directions (velocity components); the alternative notation $x_1=x$ ($u_1=u$), $x_2=y$ ($u_2=v$) and $x_3=z$ ($u_3=w$) is also used.
The harmonic oscillation generates a periodic (coherent) spanwise cross-flow, that even for a turbulent streamwise flow is well described \citep{quadrio-sibilla-2000} by the analytical laminar solution of the second Stokes problem, usually referred to as the Stokes layer:
\begin{equation}
w(y,\varphi) = A \exp \left( - \sqrt{ \frac{\omega}{2\nu} } y \right) \sin \left( \varphi - \sqrt{ \frac{ \omega }{2\nu} } y \right),
\end{equation}
where $\varphi$ is the phase of the oscillation, and $\omega=2\pi/T$. 
Figure \ref{fig:channel} shows the coherent spanwise velocity field (the Stokes layer) generated by the harmonic oscillations, and its derivative in wall-normal direction (the Stokes shear): the oscillating period is subdivided into eight equally spaced phases $\varphi_1, \varphi_2, \ldots \varphi_8$, where $\varphi_i = i \pi/4$. 
From here on, the $+$ superscript is used to indicate quantities made dimensionless with the friction velocity $u_\tau = \sqrt{\tau_w/\rho}$ ($\rho$ is the fluid density, and $\tau_w$ is the time-averaged streamwise wall shear stress; the spanwise component is zero) and the kinematic viscosity $\nu$. 

The interaction between the coherent Stokes layer and the stochastic near-wall turbulence influences the main structures of the near-wall cycle, i.e. the low-speed streaks and the quasi-streamwise vortices, eventually yielding a reduction of turbulent friction. 
When the Reynolds number based on the friction velocity is $Re_\tau=200$, the largest drag reduction rate for a given oscillation amplitude $A^+=12$ is about $45\%$, obtained for the optimal actuation period $T^+ \approx 100$ \citep{quadrio-ricco-2004}. Larger or smaller periods result in smaller drag reduction.
Several authors, for example \cite{yakeno-hasegawa-kasagi-2014}, observed that the orientation of near-wall structures in wall-parallel planes is cyclically altered by the coherent spanwise shear.
\cite{touber-leschziner-2012} have shown that, provided the timescale of the spanwise shear oscillation is short enough, the low-speed streaks do not have the time to fully re-orient during the oscillation, and are thus weakened.
Hence, at the root of drag reduction lies the interaction between the oscillating shear (a coherent component) and the natural streak regeneration mechanism (seen in the stochastic component). 

\cite{touber-leschziner-2012} and later \cite{agostini-touber-leschziner-2014} applied a triple decomposition of the velocity field to the budgets of the single-point Reynolds stresses; the turbulent (stochastic) fluctuations were isolated and their interaction with the (coherent) Stokes layer was studied. 
It was found that the interaction between coherent and stochastic fields is mediated by the interplay between the coherent spanwise shear $\partial \coh{w}/ \partial y$ and the $\paver{v'' w''}$ component of the Reynolds stress tensor, induced by the rotation of the vortical structures.
For nearly optimal periods, the interaction between the coherent and stochastic fields is one-way, with the former altering the latter. This weakens the wall-normal velocity fluctuations and reduces the turbulent shear, reducing eventually the friction drag. 
For larger periods, instead, the interaction becomes two-ways, with coherent and stochastic fields mutually exchanging energy. In this case, however, the drag reduction effect is less important. 
By looking at different phases along the period, they found that, when large, the Stokes shear $\partial \coh{w}/ \partial y$ changes relatively slowly in time and allows the structures to become more vigorous and well-established (a process they referred to as lingering). 
Conversely, when $\partial \coh{w}/ \partial y$ is small, the structures appear weak and less tilted.

In this example, we intend to add scale information to the picture. We thus apply the $\varphi$AGKE: (i) to describe the influence of the coherent motion on the spatial arrangement of the near-wall structures during the control period, (ii) to inspect the mean-coherent-stochastic interaction in the scale space and in the physical space, and (iii) to characterise the phase dependence of the interaction between the coherent and stochastic fields.

%----------------------------------------------
\subsection{Database and computational details}
\label{sec:computationaldetails}

The $\varphi$AGKE terms are computed from two datasets obtained by direct numerical simulations (DNS). They are described by \cite{gallorini-quadrio-gatti-2022}, where the interested reader can find full details. 

The simulations are run under a constant pressure gradient (CPG) \citep{quadrio-frohnapfel-hasegawa-2016}, with a friction Reynolds number of $Re_\tau = u_\tau h / \nu =200$, where $h$ is the channel half-height. CPG provides a unique value of $u_\tau$ with/without drag reduction, thus avoiding ambiguities in viscous scaling. 
The size of the computational domain is $(L_x,L_y,L_z) = (4\pi h, 2h, 2\pi h)$ in the streamwise, wall-normal and spanwise directions. 
The number of Fourier modes is $N_x= N_z= 256$ in the two homogeneous (streamwise and spanwise) directions, further increased by a factor of $3/2$ to remove aliasing error. 
In the wall-normal direction, a hyperbolic tangent distribution of $N_y = 192$ points provides a finer grid near the wall. 
The spatial resolution is $\Delta x^+ = 6.6$ and $\Delta z^+ = 3.3$ by considering the extra modes, while $\Delta y^+$ varies from $\Delta y^+ \approx 0.5$ close to the wall to $\Delta y^+ \approx 3.7$ at the centreline.

A first simulation of a plane channel with fixed walls is run as a reference, followed by two others in which wall oscillation according to \eqref{eq:OW} is enforced. 
The oscillation amplitude is fixed at $A^+=7$: a rather small value, which keeps the energy cost of the actuation limited, and might even provide a small net energy saving at optimal periods. 
As in \cite{agostini-leschziner-2014}, we consider two control periods, namely $T^+=75$ and $T^+=250$. The value $T^+=75$ is nearly optimal, and yields drag reduction (defined here as a percentage decrease of the friction coefficient, determined by the increase in bulk velocity) of 25.2\%. The value $T^+=250$ is suboptimal, and yields only 13.2\% drag reduction. These figures are in agreement with existing information \citep[see for example][]{gatti-quadrio-2016}.
 
Simulations are started from an uncontrolled turbulent flow field. During the initial, transient phase, the solution is advanced by setting the Courant--Friedrichs--Lewy number at $CFL=1$. 
After the transient, however, the time step is set to a fixed value, in order to synchronize data saving with predetermined control phases. 
The value of the time step is thus chosen as an integer submultiple of the forcing period that keeps the maximum $CFL$ below the unit: it is $\Delta t^+ = 0.0938$ for the smaller period, and $\Delta t^+ = 0.0781$ for the longer period. 
After the transient, $376$ complete velocity fields are saved, so that $47$ control periods are stored for later analysis, each of them divided in 8 equally spaced phases.

The $\varphi$AGKE terms are computed from the database with a post-processing code derived with modifications from that described by \cite{gatti-etal-2020}.
It employs the same important numerical optimizations described in \cite{gatti-etal-2019}, which include the computation of correlations pseudo-spectrally whenever possible. 
The code, written in the CPL computer programming language \citep{cpl-website, luchini-2021} has been validated by checking that the sum of each term of the budget of coherent and stochastic fields equals the corresponding term of $\aver{\delta u'_i \delta u'_j}$ within roundoff.
Statistical convergence of the results is verified by ensuring that the residuals of the budgets are negligible compared to the values of the production, pressure--strain and dissipation.

%----------------------------------------------------------------
\subsection{$\varphi$AGKE tailored to the channel flow with oscillating walls} 

The general form \eqref{eq:AGKEcoh} and \eqref{eq:AGKEsto} of the $\varphi$AGKE can be simplified for the problem at hand. 
Since $x$ and $z$ are homogeneous, in an indefinite plane channel the $\varphi$AGKE depend on five independent variables: the three components of the separation vector $(r_x,r_y,r_z)$, the wall-normal component of the midpoint $Y$ and the phase $\varphi$. 
Note that the finite distance between the two walls implies the constraint $r_y < 2 Y$.

In an indefinite channel flow, the $x$ direction aligns with the mean flow, hence $\vect{U}(y)=(U(y),0,0)$, and the wall-parallel derivatives of the mean velocity are zero. Moreover, in the specific case of the oscillating wall, the coherent velocity field is independent on $x$ and $z$, as the wall control law \eqref{eq:OW} is a function of time only, so that $\partial \coh{u}_i / \partial x = \partial \coh{u}_i / \partial z = 0$. Therefore, incompressibility and no-penetration at the wall dictate that the wall-normal component of the coherent field is null everywhere, i.e.  $\coh{v}(y,t)=0$.
The streamwise coherent velocity $\coh{u}$, instead, does not vanish, albeit it is known to be extremely small: \citep{yakeno-hasegawa-kasagi-2014} report it to be two orders of magnitude smaller than the spanwise coherent velocity $\coh{w}$. 
The non-zero components of the $\delta \coh{u}_i \delta \coh{u}_j$ tensor are $\delta \coh{u} \delta \coh{u}$, $\delta \coh{w} \delta \coh{w}$ and $\delta \coh{u} \delta \coh{w}$.

The specialised form of the $\varphi$AGKE for the channel flow with oscillating walls is reported in Appendix \ref{sec:tcf-phiagke}. 
It can be observed that the mean-coherent production $\mc{\pph}_{ij}$ is zero: in this particular case, there is no exchange of stresses between the mean and coherent fields, as the coherent field interacts directly with the external forcing and with the stochastic field only.
However, this term does appear in other flows, and for example is important for the flow past a bluff body \citep{alvesportela-papadakis-vassilicos-2020}, where the mean flow supports the coherent vortex shedding, which in turn supports the stochastic fluctuations. In the budget for the stochastic part, the productions $\ms{\pph}_{ij}$ and $\cs{\pph}_{ij}$ represent the two avenues for the stochastic field to interact with the mean and coherent fields, involving distinct components of $\paver{\delta u''_i \delta u''_j}$. 
The mean-stochastic production $\ms{\pph}_{ij}$ is non-zero only for $\paver{\delta u'' \delta u''}$ and for the off-diagonal components $\paver{\delta u'' \delta v''}$ and $\paver{\delta u'' \delta w''}$. 
In contrast, the coherent-stochastic production contributes to all the elements of $\paver{\delta u''_i \delta u''_j}$ except for $\paver{\delta v'' \delta v''}$, being $\cs{\pph}_{22}=0$. 

\begin{figure}
\centering
\includegraphics[width=0.7\textwidth]{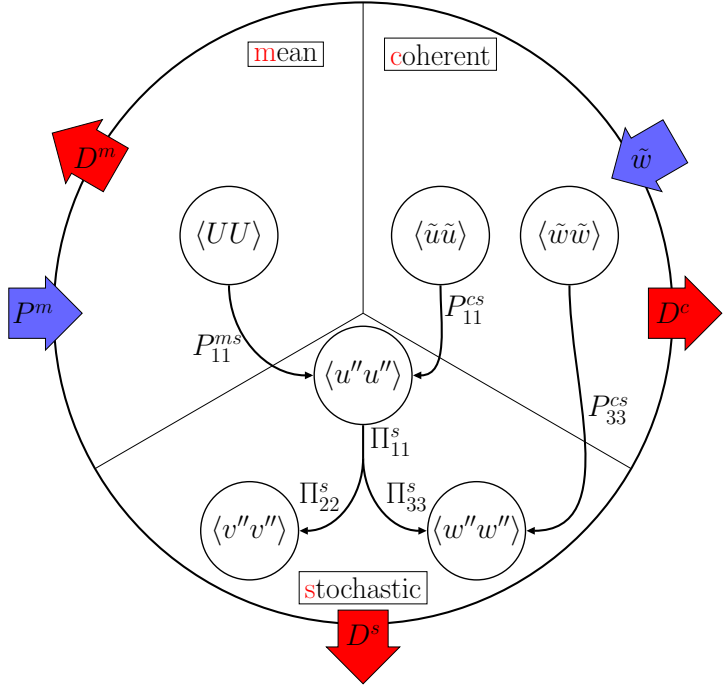}
\caption{Sketch of the energy exchanges between mean, coherent and stochastic fields for the turbulent channel flow modified by spanwise-oscillating walls. Blue/red arrows indicate energy entering/leaving the system. The blue arrows $P_m$ and $\coh{w}$ represent the pumping energy required to move the flow, and the energy introduced by the moving walls.}
\label{fig:energyBox}
\end{figure}

The flow symmetries and the type of forcing make only certain paths available for energy exchanges. This is represented graphically in figure \ref{fig:energyBox}, which shows an ``energy circle" \citep{quadrio-2011} to describe energy exchanges among the mean, coherent and stochastic fields after spatial and temporal integration. In the following, thanks to the $\varphi$AGKE, these global energy exchanges and redistributions are expanded and described in space and among scales, with a phase-by-phase analysis.

\section{Effect of the spanwise forcing on the near-wall cycle}
\label{sec:interpretation}

The influence of the oscillating wall on the structural organisation of the stochastic part of the velocity fluctuations in the near-wall region is considered first, at a single phase and then in terms of its phase evolution. 
The energy exchanges among mean, coherent and stochastic fields are then addressed, followed by the analysis of the pressure--strain redistribution. 
Eventually, the influence of the Stokes layer and the stochastic pressure--strain term $\piph^s_{33}$ on the transfer of the spanwise stochastic stresses is described.

%--------------------------------------------------------------------------
\subsection{Near-wall structures}

\subsubsection{Description at a fixed phase}
\label{sec:fixed-phase}

\begin{figure}
\centering
\includegraphics[width=1\textwidth]{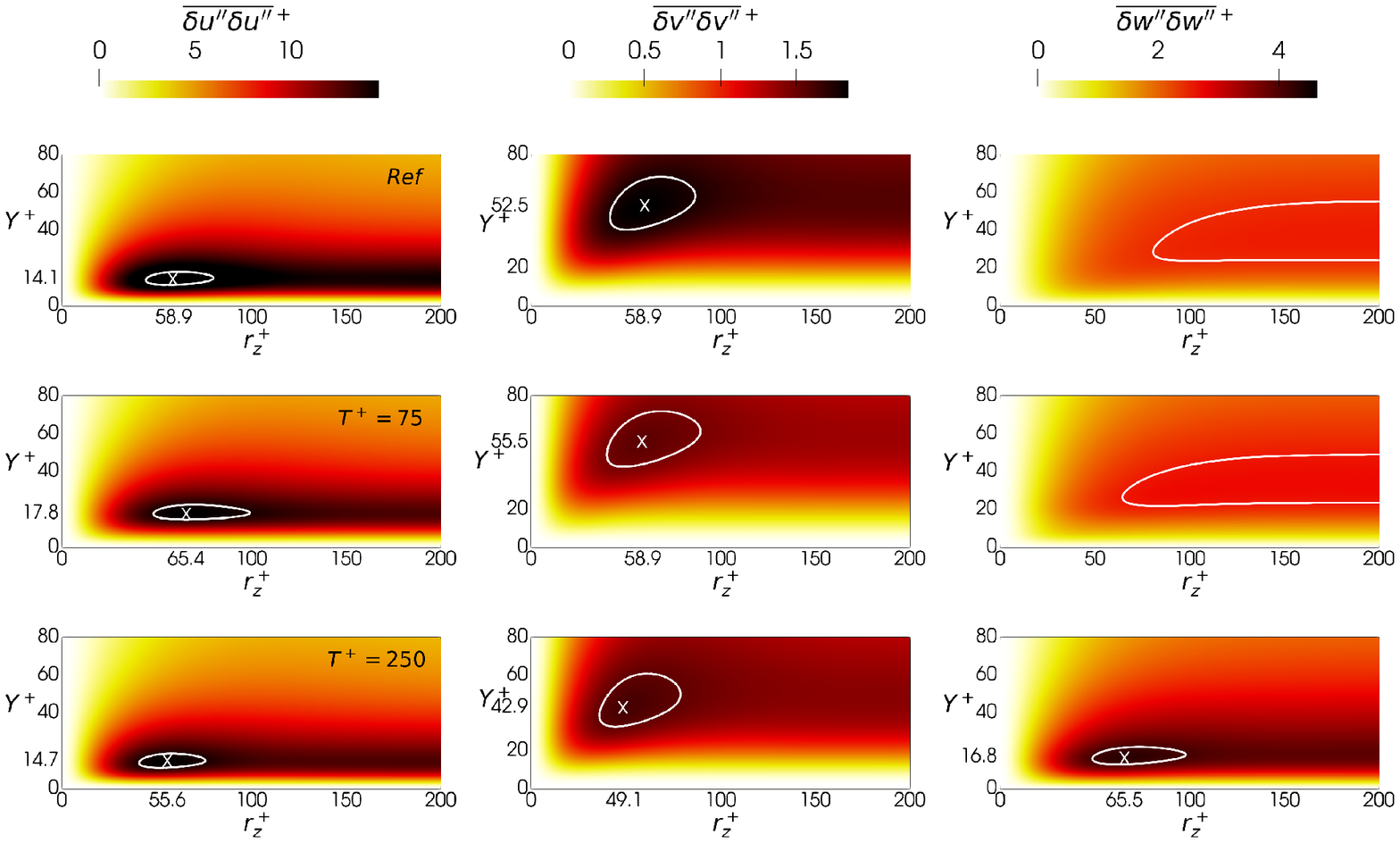}
\caption{Diagonal components of the stochastic tensor $\paver{\delta u_i'' \delta u_j''}^+$ at $\varphi_4$ in the ($r_z^+, Y^+$) plane. From top to bottom: uncontrolled case with $A=0$, $T^+=75$ and $T^+=250$. The contour is set at 95\% of each maximum. The coordinates of the maximum, marked with a cross, can be read on the axes.}
\label{fig:energy}
\end{figure}

Figure \ref{fig:energy} shows the diagonal components of $\paver{\delta u_i'' \delta u_j''}$ in the $r_y=r_x=0$ plane for the uncontrolled channel (first row), $T^+=75$ (second row) and $T^+=250$ (third row). For the two controlled cases, only phase $\varphi_4$ is shown, but the discussion that follows is qualitatively valid for all phases.

The local maxima of $\paver{\delta u'' \delta u''}$ and $\paver{\delta v'' \delta v''}$, hereafter denoted with the $\cdot_m$ subscript, are the statistical trace of the structures of the near-wall cycle. In the $r_x=r_y=0$ space, indeed, they indicate a negative peak of the streamwise and vertical stochastic correlation functions $R_{11}$ and $R_{22}$; see equation \eqref{eq:corr}.
The coordinates  $Y^+ \approx 14-18$ and $r_z^+ \approx 55-65$ of $\paver{\delta u'' \delta u''}_m$ in the $(r_z^+,Y^+)$ plane indicate the characteristic wall distance and spanwise spacing of low- and high-speed streaks.
The coordinates $Y^+ \approx 43-55$ and $r_z^+ \approx 49-59$ of $\paver{\delta v'' \delta v''}_m$ indicate the characteristic wall distance and spanwise size of the quasi-streamwise vortices, which induce at their spanwise sides regions of vertical velocity with negative correlation.

Figure \ref{fig:energy} shows that the oscillating wall leaves $\paver{\delta u'' \delta u''}$ and $\paver{\delta v'' \delta v''}$  almost unchanged, indicating that the size and strength of the near-wall structures only marginally depend on the amount of drag reduction. 

This is consistent with the CPG driving strategy, which forces the same level of wall friction; the large changes observed by various authors under different driving strategies simply derive trivially from the different friction, as discussed by  \cite{frohnapfel-hasegawa-quadrio-2012}.
However, the velocity streaks are slightly moved away from the wall: an upward shift of $\paver{\delta u'' \delta u''}_m$ can be seen in figure \ref{fig:energy}. The previous observation is confirmed by numerical data: the maximum moves from $Y^+=14.1$ in the reference case to $Y^+=17.8$ for $T^+=75$ and to $Y^+=14.7$ for $T^+=250$ (at phase $\varphi_4$). Both shifts are upwards, and the $T^+=75$ case with larger drag reduction has a larger shift.
The quasi-streamwise vortices react differently to control: $\paver{\delta v'' \delta v''}_m$ moves from $Y^+=53$ in the reference case to $Y^+=55$ for $T^+=75$ and to $Y^+=43$ for $T^+=250$. These contrasting trends are consistent with the wall-normal displacement found by \cite{gallorini-quadrio-gatti-2022} for conditionally-averaged quasi-streamwise vortices, but are extracted from the present analysis without the need for an (inevitably subjective) procedure for conditional structure extraction.

In the canonical channel flow, the map of $\paver{\delta w'' \delta w''}$ embeds information of the quasi-streamwise vortices only when the $r_y \neq 0$ space is considered, which contains the peak $\aver{\delta w' \delta w'}_m$ \citep{gatti-etal-2020}. Indeed, the quasi-streamwise vortices induce negatively correlated regions of $w''$ fluctuations at their vertical sides only, and the $r_y$ coordinate of the maximum indicates their characteristic wall-normal size. 
In the controlled cases, however, a local peak of $\paver{\delta w'' \delta w''}$ appears in the $r_x=r_y=0$ (figure \ref{fig:energy}) and $r_z=r_y=0$ (not shown) planes. Interestingly, the local peak is particularly evident for $T^+=250$, extending for $r_z^+ \approx 50-100$, $r_x^+ \approx 85-270$ and $Y^+ \approx 13-25$, but it is hardly visible for $T^+=75$, where the $w''$ fluctuations are weaker. 
The next Subsection, which examines how these quantities vary with $\varphi$, shows that this derives from a combination of the streaks tilting in the $x-z$ plane and from the interaction of the quasi-streamwise vortices with the coherent spanwise shear.

%-----------------------------------------
\subsubsection{Evolution during the cycle}
\label{sec:ph-by-ph}

\begin{figure}
\centering
\includegraphics[width=1.0\textwidth]{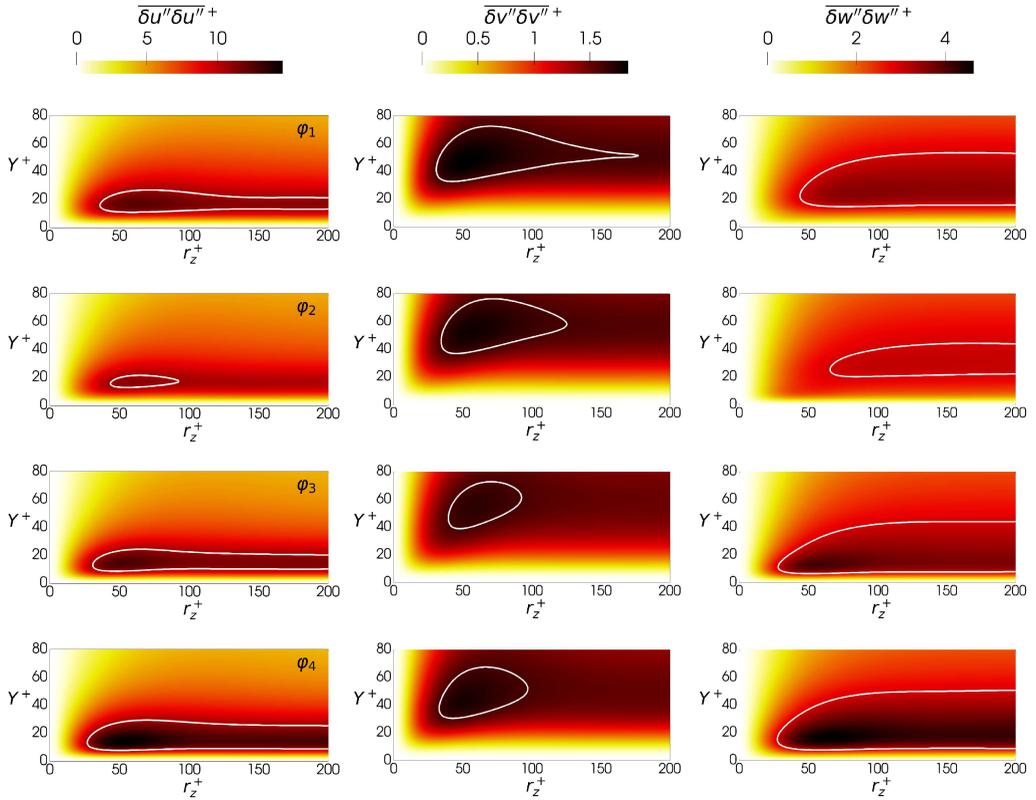}
\caption{Diagonal components of the stochastic tensor $\paver{\delta u_i'' \delta u_j''}^+$ in the $(r_z^+,Y^+)$ plane, at phases $\varphi_1, \varphi_2, \varphi_3, \varphi_4$ (from top to bottom), for the period $T^+=250$. For each component, the white contour is set at the 95\% of the smallest peak over the phases (i.e. at $\varphi_2$ for $\paver{\delta u'' \delta u''}$ and $\paver{\delta w'' \delta w''}$, and at $\varphi_3$ for $\paver{\delta v'' \delta v''}$).}
\label{fig:dudu-dvdv-dwdw-250}
\end{figure}

\begin{figure}
\centering
\includegraphics[width=1.0\textwidth]{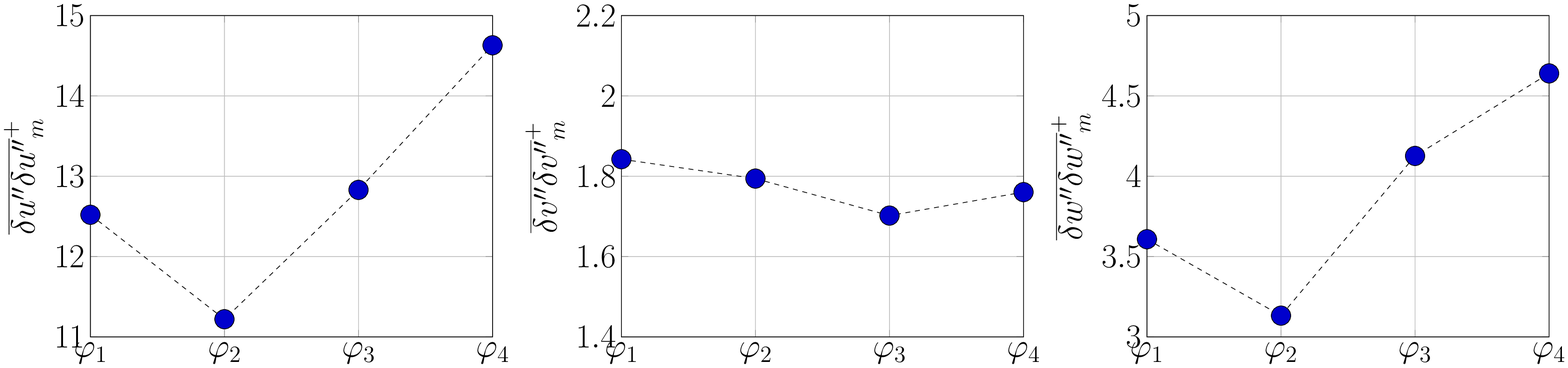}
\caption{Phase variation of the maxima $\paver{\delta u_i'' \delta u_i''}_m^+$ in the $(r_z^+,Y^+)$ plane.}
\label{fig:dudu-dvdv-dwdw-250-val}
\end{figure}

Figure \ref{fig:dudu-dvdv-dwdw-250} shows the phase evolution of $\paver{\delta u'' \delta u''}$, $\paver{\delta v'' \delta v''}$ and $\paver{\delta w'' \delta w''}$ in the $r_x=r_y=0$ plane, to describe how the organisation of the near-wall stochastic fluctuations changes during the oscillation cycle, i.e. the very type of information that the $\varphi$AGKE are designed to provide. 
Only the suboptimal $T^+=250$ is considered, as the large period emphasises the phase dependence; moreover, only one half of the forcing period is shown (from $\varphi_1$ to $\varphi_4$), because of temporal symmetry. Extra quantitative information is provided by figure \ref{fig:dudu-dvdv-dwdw-250-val}, which plots the phase evolution of the maxima $\paver{\delta u'' \delta u''}_m$, $\paver{\delta v'' \delta v''}_m$ and $\paver{\delta w'' \delta w''}_m$.

The streamwise velocity streaks cyclically strengthen and weaken under the action of the alternating Stokes layer. The maximum $\paver{\delta u'' \delta u''}_m$ assumes its lowest value at $\varphi_2$, and then grows to reach the highest value at $\varphi_4$, with an intra-cycle variation of 27\%. 
The quasi-streamwise vortices, instead, show a much smaller phase dependence: the intra-cycle variation of $\paver{\delta v'' \delta v''}$ is 8\% only. 
This is not surprising, since the quasi-streamwise vortices reside at larger wall distances, where the intensity of the Stokes layer is lower; at $y^+ = 14$, the average position of the streaks, the maximum $\coh{w}^+$ is $1.15$, while at $y^+=50$, representative wall-normal distance of the vortices, it is only $0.2$. 
A different wall distance for streaks and vortices also implies a phase shift; in fact the intensity of $\paver{\delta v'' \delta v''}$ is minimum at $\varphi_3$ and maximum at $\varphi_1$, whereas $\paver{\delta u'' \delta u''}$ and $\paver{\delta w'' \delta w''}$ are minimum at $\varphi_2$ and maximum at $\varphi_4$. This is consistent with the early observation \citep{baron-quadrio-1996} that streaks and quasi-streamwise vortices are displaced by the spanwise Stokes layer differently.

\begin{figure}
\centering
\includegraphics[width=1\textwidth]{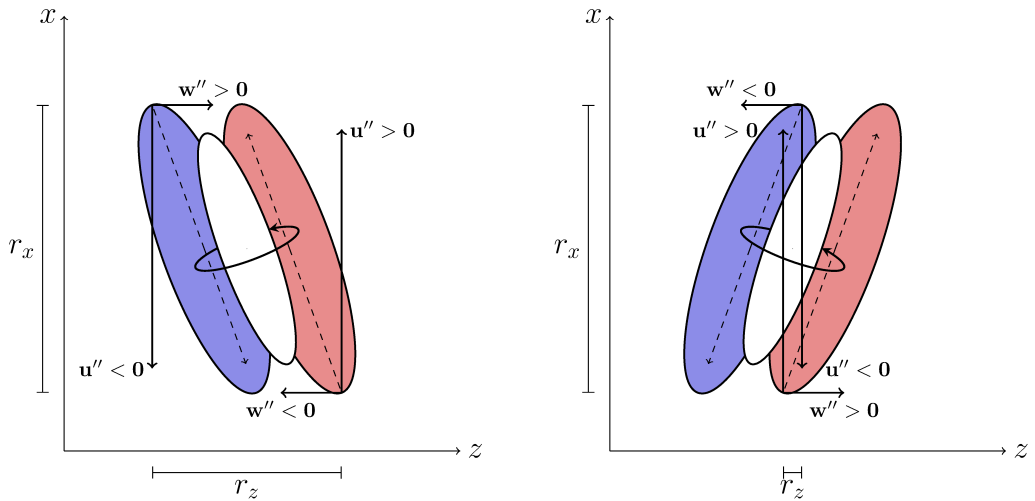}
\caption{Sketch of the contribution of $u''$ and $w''$ for positively (left) and negatively (right) tilted low (blue) and high (red) speed streaks induced by a positively rotating quasi-streamwise vortex (white).}
\label{fig:xy-sketch}
\end{figure}
From figure \ref{fig:dudu-dvdv-dwdw-250-val}, one notices that the phase evolution of $\paver{\delta w'' \delta w''}_m$ resembles that of $\paver{\delta u'' \delta u''}_m$, thus suggesting that part of the stochastic $w''$ fluctuations derives from a redistribution of the streamwise fluctuations.
The near-wall structures are tilted in the $x-z$ plane and follow the shear vector $(\text{d}U/\text{d}y,0,\partial \coh{w}/\partial y)$ \citep{yakeno-hasegawa-kasagi-2014}. The tilting causes the streamwise high- and low-speed streaks to re-orient, thus contributing via pressure--strain redistribution (see below \S\ref{sec:pstrain}) to the spanwise stochastic fluctuations. 
When the tilting angle is positive (negative), the low- and high-speed streaks contribute to respectively positive (negative) and negative (positive) $w''$. This produces regions of $w''$ fluctuations that correlate negatively for scales $r_x$ and $r_z$ and position $Y$ compatible with the position of $\paver{\delta w'' \delta w''}_m$ observed in figure \ref{fig:dudu-dvdv-dwdw-250}. 
This is shown with a sketch in figure \ref{fig:xy-sketch}, and confirmed with a phase-by-phase conditional average of events extracted from the present database in Appendix \ref{sec:ens-aver}. 
The picture is also consistent with the lower $\paver{\delta w'' \delta w''}_m$ observed in figure \ref{fig:energy} for $T^+=75$: for periods close to the optimum, the oscillation is too fast for the streaks to align with the shear vector \citep{touber-leschziner-2012}, and this redistribution mechanism becomes weaker.  

Similar information is usually extracted \citep{yakeno-hasegawa-kasagi-2014} from phase-locked conditional averages. However, such statistics are unavoidably arbitrary to some degree: e.g. ``short" structures have to be excluded from averaging, and one needs to pre-determine a specific wall distance for the eduction procedure. Here we obtain information that is equivalent to conditional averaging, but via a statistical analysis that is free from assumptions and hypotheses. 

For example, the scales $r_{z,m}$ and $r_{x,m}$ identified by $\paver{\delta w'' \delta w''}_m$ can be used to track the phase evolution of the tilting angle $\theta$ of the flow structures during the cycle:
\begin{equation}
|\theta(\varphi)|  = \tan^{-1} \left( \frac{r_{z,m}(\varphi)}{r_{x,m}(\varphi)} \right) .
\label{eq:tilting}
\end{equation}
Similarly, the evolution of the wall-normal position $Y_m$ of $\paver{\delta w'' \delta w''}_m$ (or, equivalently, of $\paver{\delta u'' \delta u''}_m$) quantifies the vertical displacement of the streaks during the cycle.
\begin{figure}
\centering
\includegraphics[width=1.0\textwidth]{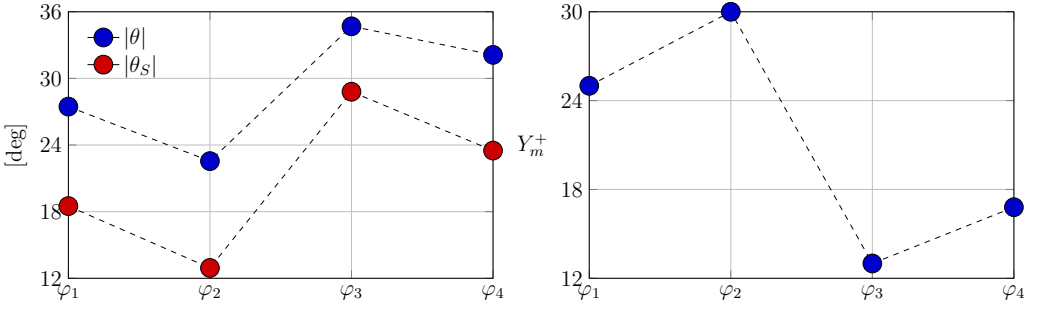}
\caption{Left: evolution of the tilt angle of the wall streaks during the cycle. Comparison between present results (blue symbols) and the shear angle introduced by \cite{yakeno-hasegawa-kasagi-2014} (red symbols). Right: wall-normal position of the structures, educed from the wall-normal position $Y_m^+$ of $\paver{\delta w'' \delta w''}_m$.}
\label{fig:angles}
\end{figure} 
Figure \ref{fig:angles} compares $|\theta|$ with the shear angle $\theta_s$ evaluated at $Y_m$, i.e.
\begin{equation*}
\theta_s =  \tan^{-1} \left( \frac{ \partial \coh{w}/\partial y}{\text{d}U /\text{d}y} \right),
\end{equation*}
that is conventionally used to describe the tilting angle of the near-wall structures \citep{yakeno-hasegawa-kasagi-2014, gallorini-quadrio-gatti-2022}. The two quantities $\theta$ and $\theta_s$ are quantitatively similar and present the same phase dependence, with a nearly constant difference of about $8^\circ$. 
The right panel of figure \ref{fig:angles} also shows that when the tilting angle of the streaks is maximum, their distance from the wall is minimum (and vice versa). This implies that a higher coherent spanwise velocity yields a larger tilting. 

Part of the wall-parallel modulation of $\paver{\delta w'' \delta w''}$ induced by the wall oscillation derives from the interaction of the quasi-streamwise vortices with the coherent spanwise shear. When the coherent shear $\partial \coh{w}/\partial y$ is positive, the quasi-streamwise vortices move low-spanwise-velocity fluid upwards, and high-spanwise-velocity fluid downwards. The opposite happens when $\partial \coh{w}/\partial y<0$.
This creates two regions with spanwise velocity of opposite sign at the vortex sides, resulting in negative $R_{33}$ correlation and a positive peak of $\paver{\delta w'' \delta w''}$ at their characteristic spanwise separation. 
This process, quantified by the coherent-stochastic production $\cs{\pph}_{33}$ (see \S\ref{sec:prod}), resembles the ejections and sweeps typical of the near-wall cycle, where the mean streamwise shear is involved; its description is similar to the explanation provided by \cite{agostini-touber-leschziner-2014} for the non-zero $\aver{v''w''}$. 
Once again, our interpretation is supported by the velocity field induced by the ensemble-averaged quasi-streamwise vortex, computed at various phases and shown in Appendix \ref{sec:ens-aver}.

%%%%%%%%%%%%%%%%%%%%%%%%%%%%%%%%%%%%%%%%%%%%%%%%%%%%%%%%%%%%%%%%%%%%
\subsection{Interaction of the mean, coherent, and fluctuating fields}
\label{sec:prod}

The energy exchanges of the mean field with the stochastic and coherent fields are described by the two mean production terms $\mc{\pph}_{ij}$ and $\ms{\pph}_{ij}$. 
However, as shown in figure \ref{fig:energyBox}, for the present problem $\mc{\pph}_{ij}=0$, and the mean field interacts directly with the stochastic field only, by feeding (or draining from) streamwise fluctuations. 
Moreover, energy is exchanged between the coherent and stochastic fields via the coherent-stochastic production $\cs{\pph}_{ij}$, which involves only $\paver{\delta u'' \delta u''}$ and $\paver{\delta w'' \delta w''}$ among the diagonal components of the $\paver{\delta u''_i \delta u''_j}$ tensor.

\begin{figure}
\centering
\includegraphics[width=1.0\textwidth]{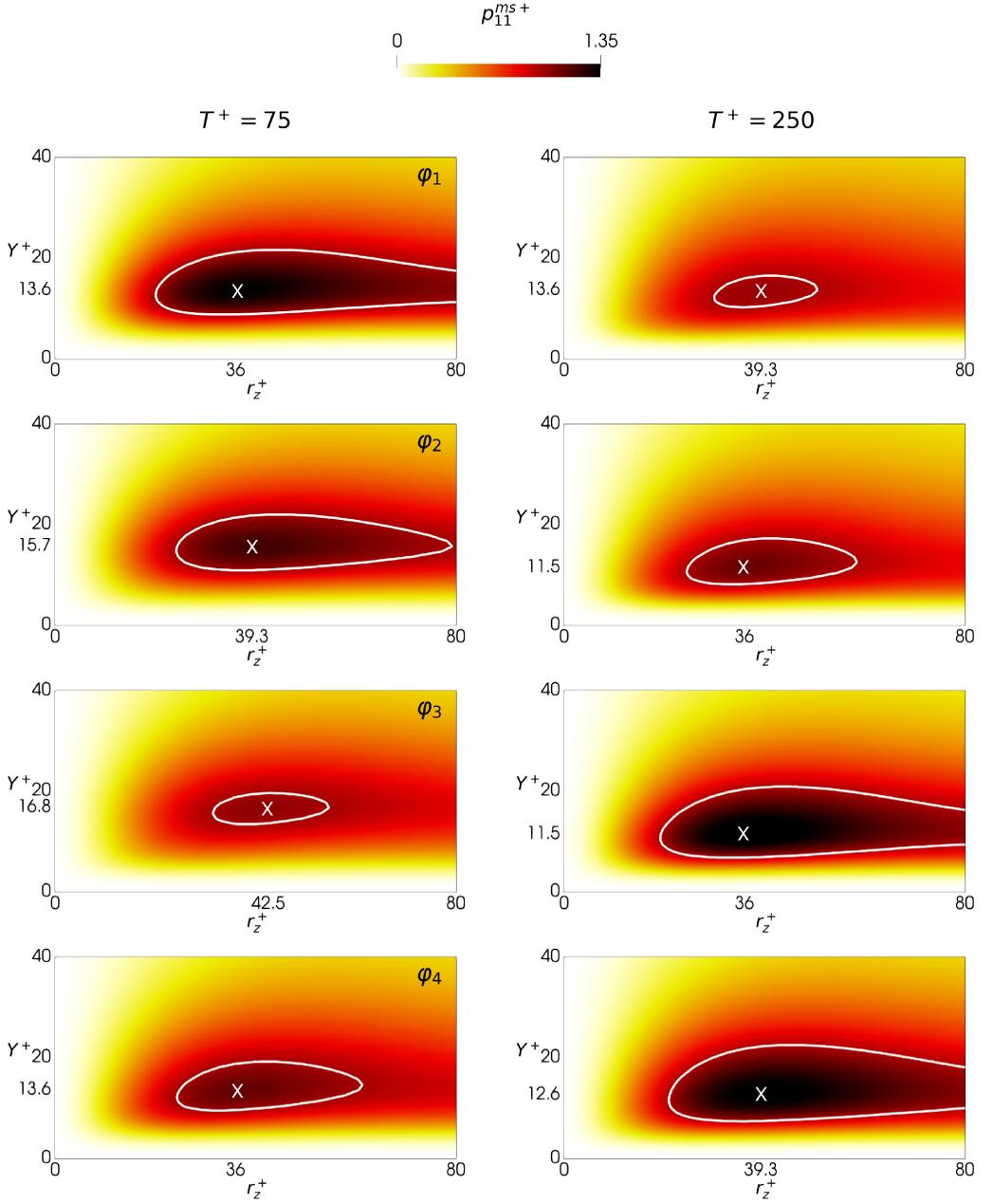}
\caption{Mean-stochastic production ${\ms{\pph}_{11}}^+$ in the $(r_z^+,Y^+)$ plane for $T^+=75$ (left) and $T^+=250$ (right). From top to bottom: $\varphi_1, \varphi_2, \varphi_3, \varphi_4$. The contour line is set at 95\% of the smallest maximum over the phases. The coordinates of the maximum, marked with a cross, can be read on the axes.}
\label{fig:pm-phases}
\end{figure}

Figure \ref{fig:pm-phases} shows how the mean-stochastic production $\ms{\pph}_{11}$ varies with $\varphi$ for $T^+=75$ (left) and $T^+=250$ (right) in the $r_x=r_y=0$ plane, where the production terms are maxima. Here $\ms{\pph}_{11}$ reduces to
\begin{equation*}
\ms{\pph}_{11} = - 2 \paver{\delta u'' \delta v''} \left( \frac{\text{d}U}{\text{d}y} \right).
\end{equation*}
The mean-stochastic production is positive everywhere, with a peak in the range $r_{z,m}^+ = 36-42$ and $Y_m^+= 13-17$ for $T^+=75$ and $r_{z,m}^+ =36-39$ and $Y_m^+ =12-14$ for $T^+=250$. Hence, the interaction of the near-wall cycle ($\paver{\delta u'' \delta v''}$) with the mean shear ($\text{d}U/\text{d}y$) invariably moves energy from the mean field towards the stochastic streamwise fluctuations. Note that the smaller $Y^+$ for $T^+=250$ is consistent with the reduced thickening of the viscous sublayer for suboptimal periods.
The production intensity is largest at $\varphi_1$ and lowest at $\varphi_3$ for $T^+=75$, whereas it is largest at $\varphi_3$ and lowest at $\varphi_1$ for $T^+=250$. Since $\text{d}U/\text{d}y$ is phase-independent, this can only descend from $\paver{\delta u'' \delta v''}$, which includes the phase evolution of the streaks and of the quasi-streamwise vortices (see \S\ref{sec:ph-by-ph} above).

\begin{figure}
\centering
\includegraphics[width=1.0\textwidth]{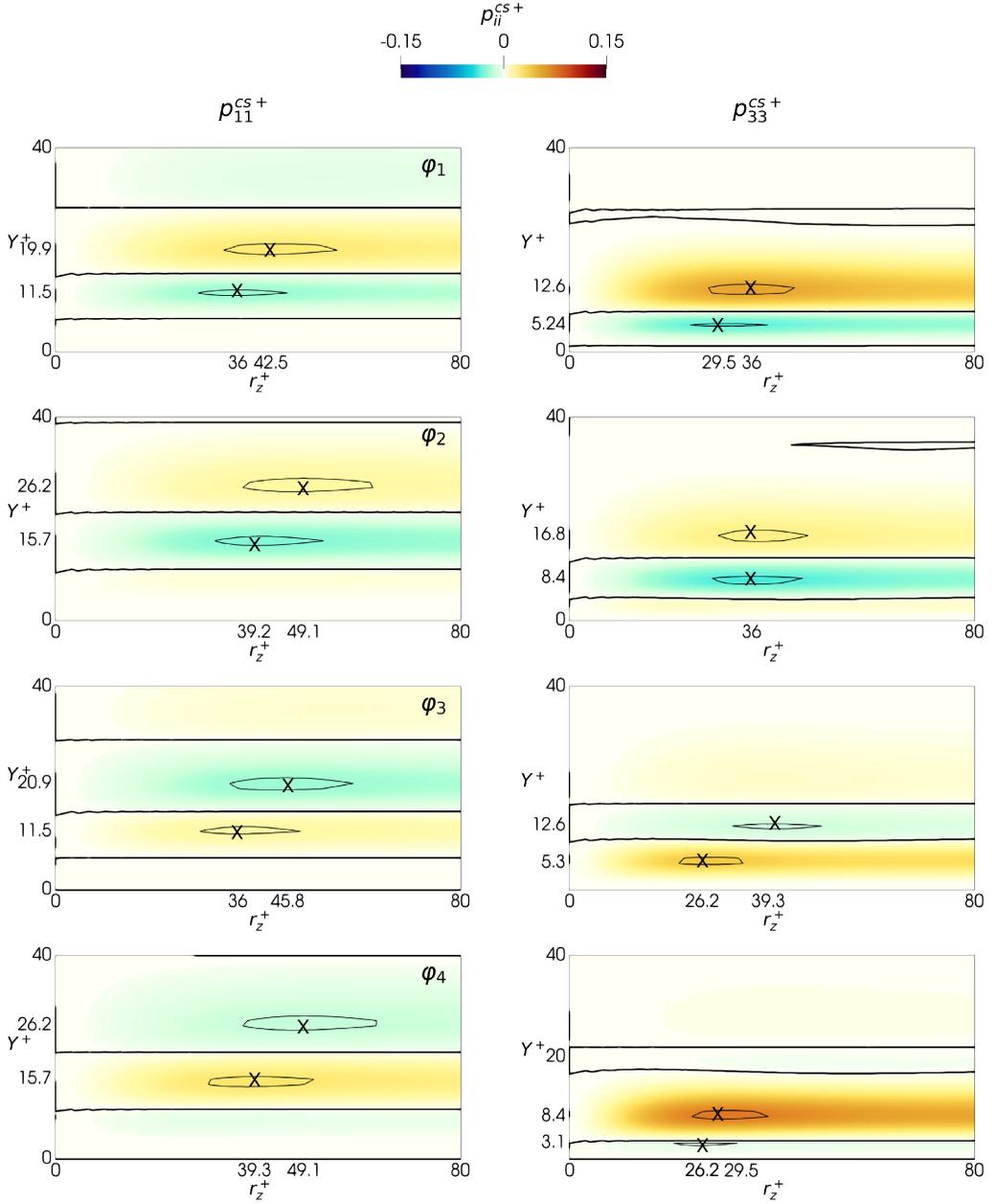}
\caption{Coherent-stochastic production ${\cs{\pph}_{11}}^+$ (left) and ${\cs{\pph}_{33}}^+$ (right) in the $(r_z^+,Y^+)$ plane for $T^+=75$. From top to bottom: $\varphi_1, \varphi_2, \varphi_3, \varphi_4$. The thin contour line is set at 95\% of the smallest (positive and negative) maximum over the phases; the thick black contour line is $\cs{\pph}_{ii}=0$. The coordinates of the maximum, marked with a cross, can be read on the axes.}
\label{fig:pc-75-phases}
\end{figure}

\begin{figure}
\centering
\includegraphics[width=1.0\textwidth]{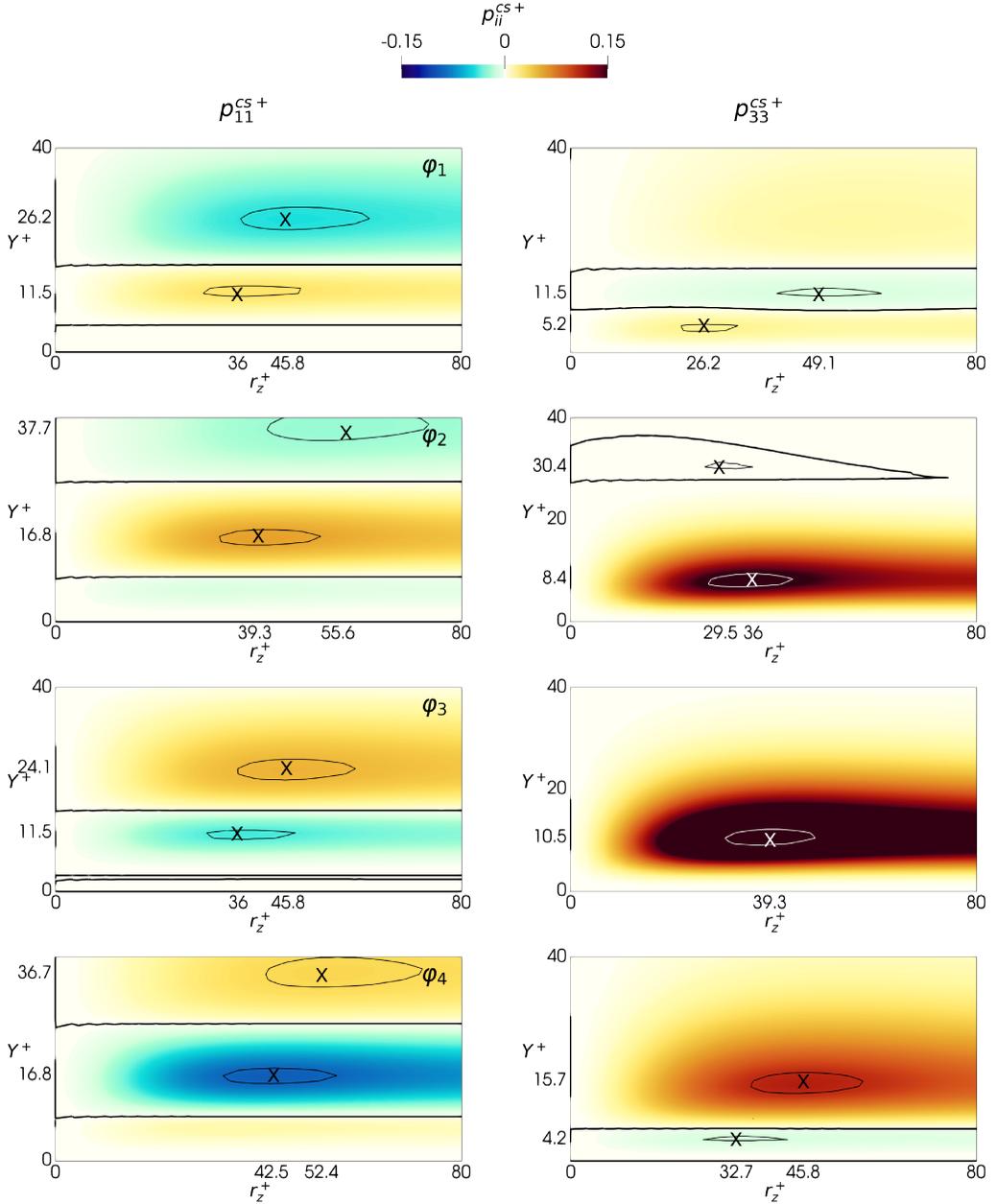}
\caption{As in figure \ref{fig:pc-75-phases}, but for $T^+=250$.}
\label{fig:pc-250-phases}
\end{figure}

Figures \ref{fig:pc-75-phases} for $T^+=75$ and \ref{fig:pc-250-phases} for $T^+=250$ show how $\cs{\pph}_{11}$ and $\cs{\pph}_{33}$ change with $\varphi$. Like for $\ms{\pph}_{11}$, the expressions for $\cs{\pph}_{11}$ and $\cs{\pph}_{33}$ simplify in the $r_x=r_y=0$ plane where their maxima occur, i.e.
\begin{equation*}
\cs{\pph}_{11} = - 2 \paver{\delta u'' \delta v''} \left( \frac{ \partial \coh{u}}{\partial y} \right)
\qquad
\text{and}
\qquad
\cs{\pph}_{33} = - 2 \paver{\delta v'' \delta w''} \left( \frac{ \partial \coh{w}}{\partial y} \right).
\end{equation*}
Unlike $\ms{\pph}_{11}$, however, these productions can take either sign. 
Their maps show evident horizontal stripes of alternating sign, from the wall up to $Y^+ \approx 40$: hence, at a given phase the coherent field feeds the stochastic field at certain wall distances, but extracts energy from it at others. It is worth noting that, although $\cs{\pph}_{11}$ and $\cs{\pph}_{33}$ at a given phase are both positive and negative, after averaging over the phases $\aver{\cs{\pph}_{11}}$ almost vanishes and $\aver{\cs{\pph}_{33}}$ is positive everywhere. 
This is not entirely new, and confirms the single-point analysis by \cite{agostini-touber-leschziner-2014} (see their figure 14); however, scale information is added here so that this mechanism can be related to the structures of the flow.
At every phase, the positive/negative peaks of $\cs{\pph}_{11}$ and $\cs{\pph}_{33}$ occur at $r_z^+ \approx 25-50$, a spanwise separation which points to the structures of the near-wall cycle. 

The intensity of $\cs{\pph}_{11}$ and $\cs{\pph}_{33}$ at the two periods is comparable, at all scales and positions. However, for $\cs{\pph}_{11}$ the contribution of the shear stresses is dominant, whereas the opposite occurs for $\cs{\pph}_{33}$, where the coherent spanwise shear dominates.
Indeed, $\partial \coh{w}/\partial y$ is two orders of magnitude larger than $\partial \coh{u}/\partial y$, and $\paver{\delta v'' \delta w''}$ is two orders of magnitude smaller than $\paver{\delta u'' \delta v''}$. Note, moreover, that for both control periods $\ms{\pph}_{11} \gg \cs{\pph}_{11}$, meaning that the streamwise stochastic fluctuations are predominantly fed by the mean field.

The alternating positive/negative stripes for $\cs{\pph}_{11}$ and $\cs{\pph}_{33}$ are due to the change of sign of $\partial \coh{u}/\partial y$ and $\partial \coh{w}/\partial y$ with $y$. For $\cs{\pph}_{33}$, the changing sign of the shear is also indirectly responsible for the alternating positive/negative $\paver{\delta v'' \delta w''}$, due to the quasi-streamwise vortices-shear interaction described in \S\ref{sec:ph-by-ph}. In contrast, for $\cs{\pph}_{11}$, $\paver{\delta u'' \delta v''}$ is entirely due to the interaction of the near-wall structures with the mean shear $\text{d}U/\text{d}y$, which overwhelms $\partial \coh{u}/\partial y$ everywhere.

Comparing figures \ref{fig:pc-75-phases} and \ref{fig:pc-250-phases} highlights that the slower oscillation introduces substantial differences in the coherent-stochastic energy exchange. 
The positive/negative maxima of $\cs{\pph}_{11}$ increase, and their position move towards larger $r_z$ and larger $Y$, but the effect of $T^+$ on $\cs{\pph}_{33}$ is even more evident. At $T^+=250$, the stripes of negative $\cs{\pph}_{33}$ weaken, while those with $\cs{\pph}_{33}>0$ strengthen: overall, the spanwise contribution to the energy flowing from the coherent to the stochastic field becomes larger.
A larger oscillating period implies a larger thickness of the Stokes layer, proportional to $\sqrt{\nu T}$, thus stretching outwards the coherent spanwise shear and, as a consequence, the scale-space map of $\paver{\delta v'' \delta w''}$, yielding an overall increase of the positive $\cs{\pph}_{33}$. 
At $\varphi_2$ and $\varphi_3$, for example, $\partial \coh{w}/\partial y$ is negative close to the wall and changes sign only at $y^+ \approx 30-50$ for $T^+=250$ (see figure \ref{fig:channel}), while it changes sign already at $y^+ \approx 13-18$ for $T^+=75$ (not shown). For $T^+=250$ this results into a large increase of the near-wall positive $\cs{\pph}_{33}$, as highlighted by the dark red colour in figure \ref{fig:pc-250-phases}.
Due to the negative $\partial \coh{w}/\partial y$, indeed, the quasi-streamwise vortices induce on their sides positive/negative $v''$ and convect upwards/downwards high/low spanwise velocity $w''$, thus yielding positive $\paver{\delta v'' \delta w''}$ and an intense energy exchange from the coherent to the stochastic field. 
The scale-space information of this exchange mechanism is highlighted by the positive peak of $\cs{\pph}_{33}$ placed at $(r_z^+,Y^+) \approx (38,9)$ for the considered $\varphi_2$ and $\varphi_3$ phases.

%----------------------------------------------------------------
\subsection{Pressure--strain redistribution}
\label{sec:pstrain}

\begin{figure}
\centering
\includegraphics[width=1.0\textwidth]{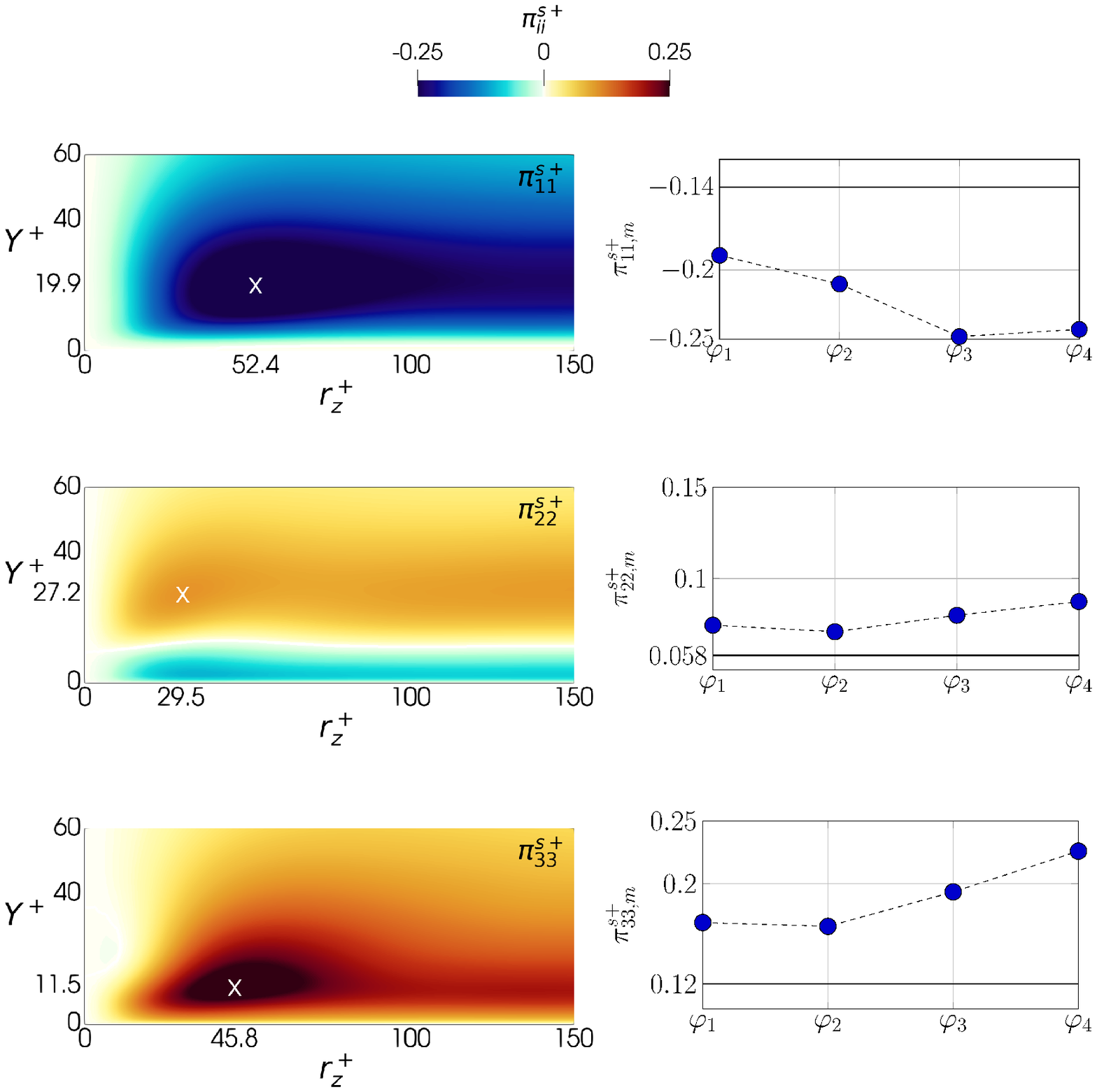}
\caption{Left: pressure--strain redistribution $\pi_{ii}^{s+}$ from $\paver{\delta u'' \delta u''}$ towards $\paver{\delta v'' \delta v''}$ and $\paver{\delta w'' \delta w''}$ at phase $\varphi_4$ for $T^+=250$; the coordinates of the maximum, marked with a cross, can be read on the axes. Right: phase variation of their maxima in the $(r_z^+,Y^+)$ plane, with a horizontal solid line indicating the value of the uncontrolled flow.}
\label{fig:pstr}
\end{figure}

As seen schematically in figure \ref{fig:energyBox}, the pressure--strain action partially redistributes the streamwise energy $\paver{\delta u'' \delta u''}$ drained from the mean flow towards the cross-stream fluctuations $\paver{\delta v'' \delta v''}$ and $\paver{\delta w'' \delta w''}$. 
The left panels of figure \ref{fig:pstr} show that $\piph_{11}^s<0$, $\piph_{22}^s>0$ and $\piph_{33}^s>0$ at almost all scales and positions: only in a very thin region close to the wall $\piph_{11}^s>0$, $\piph_{22}^s<0$ and $\piph_{33}^s>0$, according to the reorientation of vertical fluctuations into wall-parallel ones because of the impermeable wall \citep{mansour-kim-moin-1988}. 
The peaks of $\piph_{11}^s$, $\piph_{22}^s$ and $\piph_{33}^s$ in the ($r_z,Y$) plane have $Y_m^+ \approx 11-27$ and $r_{z,m}^+ \approx 30-52$, indicating that the energy redistribution is dominated by the near-wall cycle. 

It is known \citep{touber-leschziner-2012, yakeno-hasegawa-kasagi-2014} that the spanwise oscillation of the wall enhances the energy redistribution, mainly towards spanwise fluctuations. 
Compared to the uncontrolled case, the negative peak of $\piph_{11}^s$ increases by 23--67\% for $T^+=75$ and by 36--77\% for $T^+=250$, while the positive peak of $\piph_{22}^s$ decreases by 2--11\% for $T^+=75$ and increases by 4--29\% for $T^+=250$. 
The positive peak of $\piph_{33}^s$, instead, has the largest variation, with and increase of 30--53\% for $T^+=75$ and 40--87\% for $T^+=250$. 

The phase evolution of the pressure-mediated energy redistribution is described in the right panels of figure \ref{fig:pstr} for the $T^+=250$ case, by considering the maxima of the diagonal components of $\piph_{ij}^s$. Only their values are plotted, since their position remains nearly constant at $(Y^+,r_z^+) \approx (20,52)$ for $\piph^s_{11,m}$, $ \approx (27,30)$ for $\piph^s_{22,m}$ and $\approx (12,46)$ for $\piph^s_{33,m}$. 
Like $\paver{\delta v'' \delta v''}_m$, $\piph_{22,m}^s$ is the component with the smallest intra-cycle variation, with a 21\% excursion during the cycle compared to 30\% and 35\% for $\piph_{11,m}^s$ and $\piph_{33,m}^s$. 
In fact, the largest energy redistribution towards $\paver{\delta v'' \delta v''}$ occurs quite far from the wall, where the influence of the Stokes layer is weak. 
The phase dependence of $\piph_{11,m}^s$ is qualitatively different from the others. The redistribution of $\paver{\delta u'' \delta u''}$ towards the cross-stream components is maximum at $\varphi_3$ and minimum at $\varphi_1$, following the absolute value of $\piph_{11,m}^s$. In contrast, $\piph_{22,m}^s$ and $\piph_{33,m}^s$ are minima at $\varphi_2$ and maxima at $\varphi_4$ (this is not inconsistent with the incompressibility constraint $\piph_{11}^s+ \piph_{22}^s + \piph_{33}^s = 0$, since the three maxima occur at different scales and positions.) 
As already mentioned in \S\ref{sec:ph-by-ph}, $\piph_{33,m}^s$ and $\paver{\delta w'' \delta w''}_m$ have the same phase dependence, confirming that the tilting of the near-wall structures is accompanied by a redistribution of the streamwise fluctuations towards the spanwise ones.

%----------------------------------------------
\subsection{Transfers of the spanwise stresses}

A peculiarity of the present flow is the direct connection between the Stokes layer and the stochastic stresses, described by the coherent-stochastic production $\cs{P}$ shown in figure \ref{fig:energyBox}. It is therefore interesting to examine the variable-phase scale-space transfers of the stochastic stresses by looking at their fluxes in the scale and physical spaces. In this analysis, we only consider the transfer of spanwise stresses $\paver{\delta w'' \delta w''}$, since for the streamwise stresses $\cs{\pph}_{11}$ is negligible compared to $\ms{\pph}_{11}$.
Moreover, only the $T^+=250$ case is considered, as the one where the effect of the Stokes layer on the $w''$ field is larger. 
For simplicity, the analysis is restricted to the $r_x=r_y=0$ subspace, where the budget of $\paver{\delta w'' \delta w''}$ can be rewritten by moving to the r.h.s. the off-plane flux divergence terms $\partial \phiph_{x,33}^s/\partial r_x$, $\partial \phiph_{y,33}^s/\partial r_y$ and the phase evolution term, as follows:
\begin{equation}
\frac{ \partial \phiph_{z,33}^s }{ \partial r_z } + \frac{ \partial \psiph_{33}^s }{ \partial Y }
  = 
\underbrace{ \cs{\pph}_{33} + \piph_{33}^s + \dph_{33}^s}_{\xiph^s_{33}} - \frac{ \partial \phiph_{x,33}^s }{ \partial r_x } - \frac{ \partial \phiph_{y,33}^s }{ \partial r_y } - \omega \frac{\partial \paver{\delta w'' \delta w''}}{\partial \varphi} .
\end{equation}

In this way, the l.h.s. features the divergence of the in-plane flux vector, which provides information on the energetic relevance of the fluxes with its intensity, and shows their direction via its field lines. 
Moreover, the off-plane fluxes (i.e. the last three terms in the equation above) are always very small, and the in-plane divergence approximates well the full source term $\xiph_{33}^s$ everywhere \citep{gatti-etal-2020}. 
This descends from a combination of the symmetries owned by the plane channel flow system, and of the approximate alignment of the dominant vortical structures with the streamwise direction. 
Hence, the scale-space properties of the source term $\xi_{33}^s$ approximate well those of the divergence of the in-plane flux.

\begin{figure}
\centering
\includegraphics[width=1.0\textwidth]{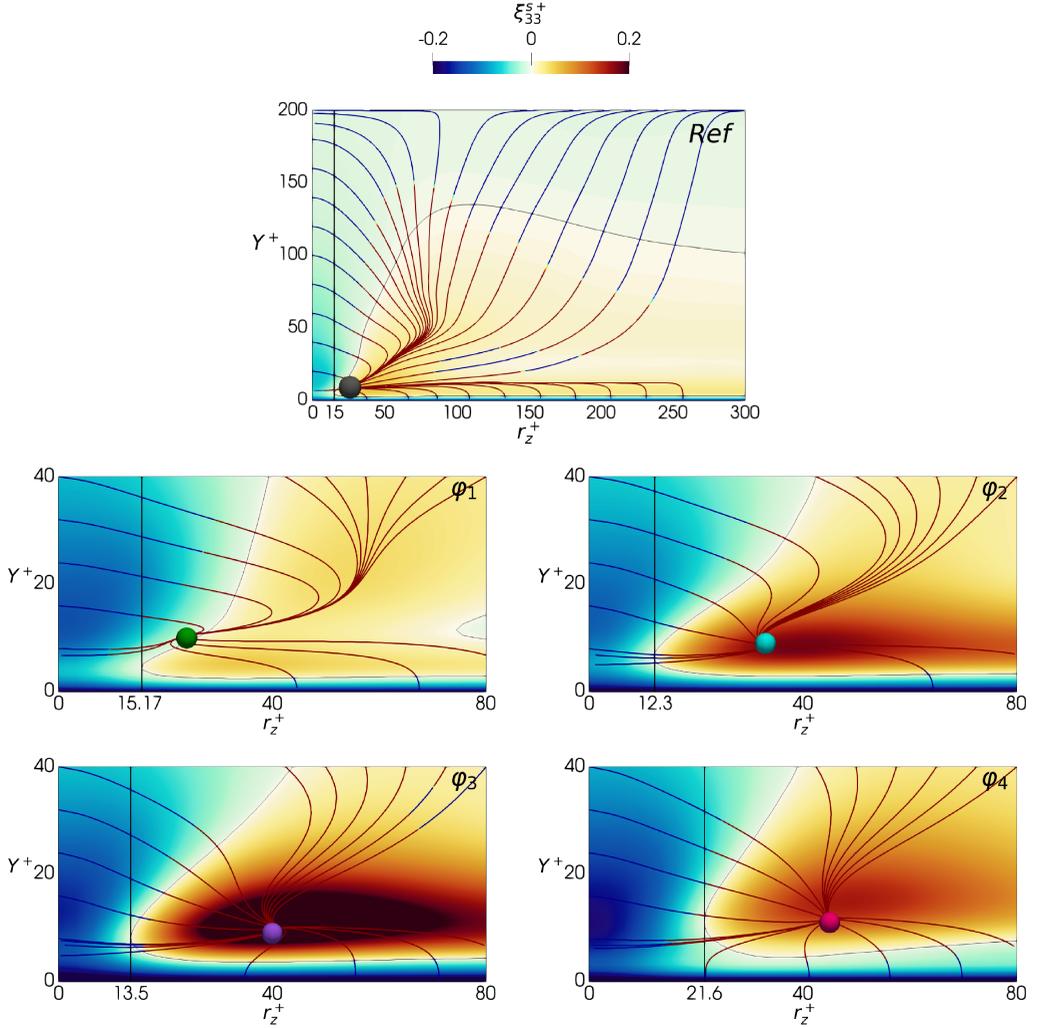}
\caption{Source $\xiph_{33}^{s+}$ in the $(r_z^+,Y^+)$ plane, with field lines of the in-plane flux vector coloured with its divergence for the uncontrolled case (top) and for the controlled case at $T^+=250$ at phases $\varphi_1$, $\varphi_2$, $\varphi_3$ and $\varphi_4$. The thin contour line marks the zero level. Dots (coloured according to figure \ref{fig:channel}) indicate the singularity point for the near-wall source, and the black vertical line marks the cut-off spanwise scale $r_{z,min}^+$ (see text).}
\label{fig:wwFluxes}
\end{figure}

\begin{figure}
\centering
\includegraphics[width=1.0\textwidth]{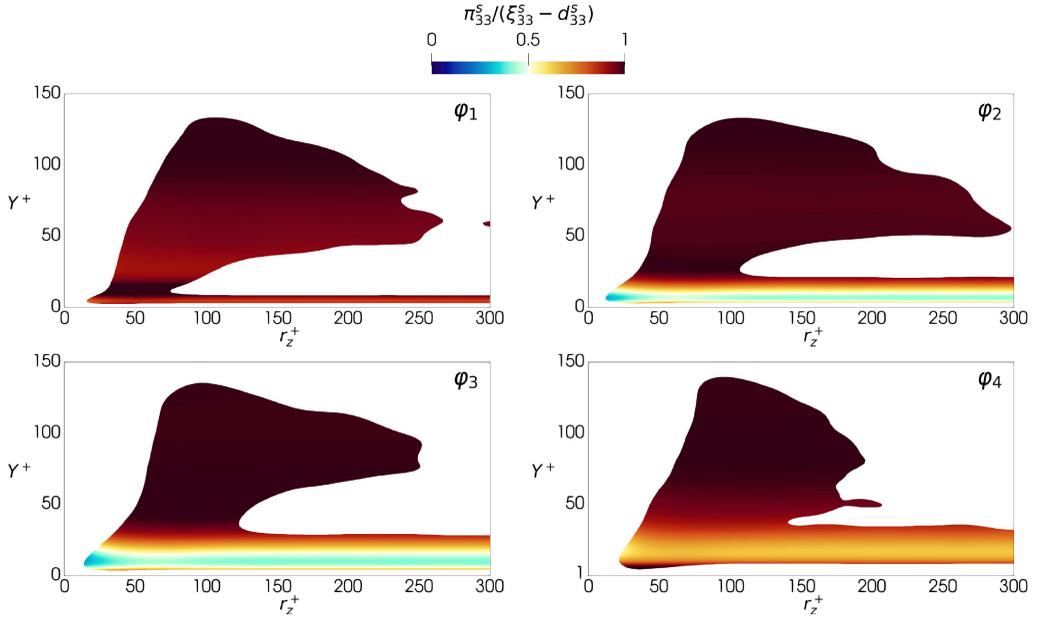}
\caption{Region with positive source in the $(r_z^+,Y^+)$ plane at phases $\varphi_1$, $\varphi_2$, $\varphi_3$ and $\varphi_4$ for $T^+=250$. The colour scale is for the ratio $\piph_{33}^s/( \xiph_{33}^s - \dph_{33}^s )$ and is centered at 0.5: red means $\piph_{33}^s>\cs{\pph}_{33}$, and blue means $\piph_{33}^s<\cs{\pph}_{33}$.}
\label{fig:PstrDivPc}
\end{figure}

Figure \ref{fig:wwFluxes} plots the map of $\xiph_{33}^s= \cs{\pph}_{33} +\piph_{33}^s + \dph_{33}^s$ for the uncontrolled case (where $\cs{\pph}_{33}=0$) and the controlled case at $T^+=250$ for $\varphi_1$, $\varphi_2$, $\varphi_3$ and $\varphi_4$, with the field lines of the in-plane flux coloured with its divergence.
In the uncontrolled case, a region with $\xiph^s_{33}>0$ extends for $5 \lessapprox Y^+ \lessapprox 100$ and for $r_z^+ \gtrapprox 15$, at scales and positions where the pressure--strain dominates over dissipation. When control is active, instead, $\xiph_{33}^s$ receives the additional contribution from coherent-stochastic production, and the values of $\xiph^s_{33}$ are generally larger. 
Two regions with $\xiph^s_{33}>0$ exist. One is close to the wall at $Y^+ \approx 10-20$, and extends for all scales $r_z^+ \gtrapprox 15$, with a peak at $r_z^+ \approx 40$. A second, connected region involves larger wall distances and scales, in the $40 \lessapprox r_z^+ \lessapprox 200$ range. It is clearly visible in figure \ref{fig:PstrDivPc}, where the ratio $\piph_{33}^s/( \xiph_{33}^s - \dph_{33}^s )$ is plotted to determine the main contribution to these positive sources at the different phases. When $\piph_{33}^s/( \xiph_{33}^s - \dph_{33}^s )> 0.5$, $\piph_{33}^s> \cs{\pph}_{33}$ meaning that the pressure--strain is the largest contribution to the positive source.
When $\piph_{33}^s/( \xiph_{33}^s - \dph_{33}^s )< 0.5$, instead, the main contributor is the coherent-stochastic production $\cs{\pph}_{33}$. 
Figure \ref{fig:PstrDivPc} shows that $\cs{\pph}_{33}$ and $\piph^s_{33}$ contribute both to the near-wall source, but their relative importance changes with the phase. 
For $\varphi_2$ and $\varphi_3$ $\cs{\pph}_{33}$ is the main contributor to the intense source peak. 
For $\varphi_1$ and $\varphi_4$, instead, $\cs{\pph}_{33}$ weakens (see figure \ref{fig:pc-250-phases}): now the (weaker) source is mainly fed by the pressure--strain. 
The source at larger $Y$, instead, is dominated by the pressure--strain at all phases; this is reasonable, as for $y^+ > 30$ the Stokes layer and consequently the coherent-stochastic production are weak.

As for the sinks, figure \ref{fig:wwFluxes} shows three of them: viscous effects dominate the very near-wall region ($Y \rightarrow 0$), the bulk flow ($Y \rightarrow h$), and the smallest scales ($r_z \rightarrow 0$).
Extension and intensity of these sinks change with $\varphi$, according to the evolution of $\cs{\pph}_{33}$, $\piph_{33}^s$ and $\dph_{33}^s$. A cut-off scale $r_{z,min}$ \citep{chiarini-etal-2022} can also be plotted to quantify the minimal scale where (spanwise) energy is always dissipated, regardless of the wall distance.

The field lines of $\paver{\delta w'' \delta w''}$ drawn in figure \ref{fig:wwFluxes} originate from a singularity point, i.e. a point near the source peak where the direction of the fluxes is undefined.
Here the lines are energised by the intense positive source and transfer $\paver{\delta w'' \delta w''}$ towards the sinks.
Three types of lines are recognised, depending on where they vanish, and reflect the three sinks described above. 
Overall, these fluxes indicate the coexistence of ascending/descending and direct/inverse energy transfers, as described by \cite{cimarelli-deangelis-casciola-2013, cimarelli-etal-2016, chiarini-etal-2021} in the context of Poiseuille and Couette turbulent flows. 

The three line types possess the same topology in the uncontrolled and controlled cases. 
For the latter, though, the amount of spanwise energy withdrawn from the sources and released to the sinks changes with $\varphi$. 
An estimate of this change is provided by the phase evolution of the positive peak of the two-dimensional divergence of the flux vector. Its value is maximum at $\varphi_3$ where it is $3.36$, $1.56$ and $1.29$ times larger than at $\varphi_1$, $\varphi_2$ and $\varphi_4$ respectively. This is consistent with the phase evolution of the positive peak of $\cs{\pph}_{33}$ visualised in figure \ref{fig:pc-250-phases}.
Moreover, the singularity point lies in the source region dominated by $\cs{\pph}_{33}$, and its $r_z$ position moves with $\varphi$ following the peak of $\cs{\pph}_{33}$, being $r_z^+=24$, $33$, $40$ and $45$ for $\varphi_1$, $\varphi_2$, $\varphi_3$ and $\varphi_4$; for the uncontrolled case it is $r_z^+=26$.

We therefore conclude that, at least for the $T^+=250$ case discussed here, the phase dependence of the transfers of $\paver{\delta w'' \delta w''}$ is governed by the $\cs{\pph}_{33}$ contribution to $\xiph_{33}^s$ rather than by $\piph_{33}^s$.
At all phases, the largest part of the $\paver{\delta w'' \delta w''}$ withdrawn by the source is released in the near-wall region; a relatively smaller part goes to the smallest scales, and a minimal part goes towards the channel centre, where the turbulent activity is low.
By comparing the negative peaks of the divergence of the in-plane flux vector at the wall and at the smallest scales, it is established that in the uncontrolled case the amount of $\paver{\delta w'' \delta w''}$ released at $Y \rightarrow 0$ is $5.67$ times larger than that released at $r_z \rightarrow 0$. The oscillating wall alters the relative importance of the fluxes: the amount of $\paver{\delta w'' \delta w''}$ released at $Y \rightarrow 0$ is significantly less, being  $2.62$, $3.85$, $2.46$ and $2.41$ times larger than that released at $r_z \rightarrow 0$, at phases $\varphi_1$, $\varphi_2$, $\varphi_3$ and $\varphi_4$ respectively.

\section{Concluding discussion}
\label{sec:conclusions}

We have derived the phase-aware anisotropic generalised Kolmogorov equations or $\varphi$AGKE, inferred from the incompressible Navier--Stokes equations, after a triple decomposition to separate the velocity and pressure fields into their coherent and stochastic parts. 

The $\varphi$AGKE are exact budget equations for the coherent and stochastic contributions to the second-order structure function tensor, namely $\delta \coh{u}_i \delta \coh{u}_j(\vect{X},\vect{r},\varphi)$ and $\paver{\delta u''_i \delta u''_j}(\vect{X},\vect{r},\varphi)$. 
Compared to the standard AGKE, which are based on the classic (double) Reynolds decomposition, the $\varphi$AGKE add extra features. 
(i) The transport equations for the coherent and stochastic parts are separated: disentangling their dynamics becomes possible. 
(ii) The scale-space energy exchanges among mean, coherent, and stochastic fields can be tracked. 
In particular, the mean-coherent production $\mc{\pph}_{ij}$ and the mean-stochastic production $\ms{\pph}_{ij}$ bring out scales and positions where the mean flow feeds, and/or drains energy from, the coherent and stochastic fields; the coherent-stochastic production $\cs{\pph}_{ij}$ describes the exchange between the coherent and stochastic fields. 
(iii) An extra term in the budget for $\delta \coh{u}_i \delta \coh{u}_j$ represents the mutual interaction of the coherent motions at different phases. 
(iv) The $\varphi$AGKE imply no average over phases, and thus describe the phase variation of the various terms related to coherent and stochastic motions. 
Once a phase average is taken, as in \cite{alvesportela-papadakis-vassilicos-2020}, phase information is obviously lost.

To demonstrate the potential of the $\varphi$AGKE, we have applied them to a turbulent plane channel flow in which spanwise wall oscillations reduce the turbulent skin friction. 
The $\varphi$AGKE are perfectly suited for this flow, owing to its deterministic and periodic external forcing; moreover, the physics of drag reduction remains not entirely understood and contains interesting inter-phase and multi-scale dynamics.

Thanks to the $\varphi$AGKE, the phase-dependent modifications of the near-wall turbulent structures have been observed without the need for somewhat arbitrary procedures to educe phase-locked and conditionally-averaged structures. The flow scales involved in the redistribution of fluctuating energy have been described, together with the process by which streamwise velocity fluctuations are converted into spanwise ones by the action of pressure--strain.
The interaction among the mean, coherent, and stochastic fields is easily observed with the $\varphi$AGKE, which highlight the energy exchanges between the coherent and stochastic fields, driven by the interaction between the quasi-streamwise vortices and the coherent spanwise shear. 
The phase-by-phase, scale-space transfers of the spanwise stochastic stresses, observed here for the first time, have revealed a significant phase dependency for the spanwise energy fluxes, which present ascending/descending and direct/inverse energy transfers at all phases.

The $\varphi$AGKE can be leveraged to arrive at a thorough description of two-points second-order statistics in cases that reach far beyond the oscillating-wall problem, used here as a representative example only. 
Turbulent flows where an external periodic forcing is present are common: oscillating airfoils, rotors and turbines are only a few examples.
Moreover, the $\varphi$AGKE can also be used to tackle turbulent flows without a strictly periodic forcing, in which stochastic fluctuations coexist with some kind of coherent motion. 
A non-exhaustive list includes the turbulent flow past bluff bodies, where large-scale motions typical of the K\'arm\'an-like vortices in the wake coexist with the stochastic motion of smaller scale \citep{provansal-mathis-boyer-1987}; the Taylor--Couette flow, in which Taylor--G\"ortler vortices develop and remain visible well into the turbulent regime \citep{koschmieder-1979}; the atmospheric boundary layer, rich with quasi-two-dimensional structures forced at smaller scales \citep{young-etal-2002}.  
In such cases, though, the period of the oscillation is not uniquely identified, and attention has to be paid to properly define a phase reference.

Lastly, it should be realised that the specific triple decomposition behind the $\varphi$AGKE does not matter: alternatives to the temporal triple decomposition could be used with a different meaning attached to the $\tilde{\cdot}$ and $\cdot''$ operators, without altering the ensuing equations. One example is the spatial triple decomposition approach adopted for example by \cite{bech-andersson-1996} and \cite{gai-etal-2016} to decompose the velocity fluctuations into secondary flow and residual fluctuations in a rotating turbulent plane Couette flow.
A further use case for the $\varphi$AGKE would be a turbulent flow over a flat wall with a periodic pattern, like e.g. riblets or dimples, in which the phase average would be again spatially defined.  
Finally, another option is to employ a scale-based triple decomposition. For example, \cite{andreolli-quadrio-gatti-2021} used a scale decomposition mutuated from \cite{kawata-alfredsson-2018} to separate the fluctuating velocity field in a Couette flow into small- and large-scale components, examining the kinetic energy budget of both components in physical space. 
This information, compacted by \cite{andreolli-quadrio-gatti-2021} through spatial integration into an energy budget without independent variables, similar to that in figure \ref{fig:energyBox}, can instead be expanded at will in the full physical and scale space thanks to the $\varphi$AGKE, thus providing the ultimate information about two-points second-order statistics of the flow.

\appendix
\section{Derivation of the budget equations for $\delta \coh{u}_i \delta \coh{u}_j$ and $\paver{\delta u''_i \delta u''_j}$}
\label{sec:eqderiv}

The derivation of the $\varphi$AGKE equations via triple decomposition is described below, by listing the sequence of the main analytical steps.

%--------------------------------------------------------------
\subsection{Budget equation for $U_i$, $\coh{u}_i$ and $u''_i$}

The starting point is the incompressible Navier--Stokes equations:
\begin{equation}
\frac{\partial u_i}{\partial t} + u_k \frac{\partial u_i}{\partial x_k} = - \frac{1}{\rho} \frac{\partial p}{\partial x_i} + \nu \frac{\partial^2 u_i}{\partial x_k \partial x_k} + f_i.
\end{equation}
The triple decomposition \eqref{eq:triple_decomposition} for $u_i$, $p$ and $f_i$ is introduced to obtain:
\begin{equation}
\begin{gathered}
\frac{\partial \coh{u}_i}{\partial t} + \frac{\partial u''_i}{\partial t} + \left( U_k + \coh{u}_k + u''_k \right) \frac{\partial}{\partial x_k} \left( U_i + \coh{u}_i + u''_i \right) = - \frac{1}{\rho} \frac{\partial}{\partial x_i} \left( P + \coh{p} + p'' \right) +\\ + \nu \frac{\partial^2}{\partial x_k \partial x_k} \left( U_i + \coh{u}_i + u''_i \right) + F_i + \coh{f}_i + f''_i 
\end{gathered}
\end{equation}
which can be reorganised as
\begin{equation}
\begin{gathered}
\frac{\partial \coh{u}_i}{\partial t} + \frac{\partial u''_i}{\partial t} + 
    U_k   \frac{\partial U_i}{\partial x_k} +      U_k  \frac{\partial \coh{u}_i}{\partial x_k} +      U_k  \frac{\partial u''_i}{\partial x_k} + 
\coh{u}_k \frac{\partial U_i}{\partial x_k} + \coh{u}_k \frac{\partial \coh{u}_i}{\partial x_k} + \coh{u}_k \frac{\partial u''_i}{\partial x_k} + \\ +
    u''_k   \frac{\partial U_i}{\partial x_k} +      u''_k  \frac{\partial \coh{u}_i}{\partial x_k} +      u''_k  \frac{\partial u''_i}{\partial x_k} = 
    - \frac{1}{\rho} \frac{\partial P}{\partial x_i} - \frac{1}{\rho} \frac{\partial \coh{p}}{\partial x_i} - \frac{1}{\rho} \frac{\partial p''}{\partial x_i} + \\ +
    \nu \frac{\partial^2 U_i}{\partial x_k \partial x_k} + \nu \frac{\partial^2 \coh{u}_i}{\partial x_k \partial x_k} + \nu \frac{\partial^2 u''_i}{\partial x_k \partial x_k} + F_i + \coh{f}_i + f''_i.
\end{gathered}
\label{eq:Ns-3decomp}
\end{equation}
Now the averaging operator $\aver{\cdot}$ is used to arrive at the budget equation for $U_i$, i.e.
\begin{equation}
\begin{gathered}
U_k \frac{\partial U_i}{\partial x_k} + \aver{\coh{u}_k \frac{\partial \coh{u}_i}{\partial x_k}} + \aver{u''_k \frac{\partial u''_i}{\partial x_k}} =
-\frac{1}{\rho} \frac{\partial P}{\partial x_i} + \nu \frac{\partial^2 U_i}{\partial x_k \partial x_k} + F_i.
\end{gathered}
\label{eq:MeanUBudget}
\end{equation}
When, instead, the phase average operator $\paver{\cdot}$ is used, we get:
\begin{equation}
\begin{gathered}
\frac{\partial \coh{u}_i}{\partial t} +
 U_k \frac{\partial      U_i }{\partial x_k} + \coh{u}_k \frac{\partial      U_i }{\partial x_k} +
 U_k \frac{\partial \coh{u}_i}{\partial x_k} + \coh{u}_k \frac{\partial \coh{u}_i}{\partial x_k} +
 \paver{ u''_k \frac{\partial u''_i}{\partial x_k} } = \\ - \frac{1}{\rho} \frac{\partial P}{\partial x_i} - \frac{1}{\rho} \frac{\partial \coh{p}}{ \partial x_i}  + 
 \nu \frac{\partial^2 U_i}{\partial x_k \partial x_k} + \nu \frac{\partial ^2 \coh{u}_i}{\partial x_k \partial x_k} + F_i + \coh{f}_i 
\end{gathered}
\end{equation}
which can be written differently using the budget equation for $U_i$, i.e.:
\begin{equation}
\begin{gathered}
\frac{\partial \coh{u}_i}{\partial t} + \coh{u}_k \frac{\partial U_i}{\partial x_k} + U_k \frac{\partial \coh{u}_i}{\partial x_k} + \coh{u}_k \frac{\partial \coh{u}_i}{\partial x_k} +\paver{ u''_k \frac{\partial u''_i}{\partial x_k}} -\aver{ \coh{u}_k \frac{\partial \coh{u}_i}{\partial x_k}} - \aver{      u''_k  \frac{\partial u''_i      }{\partial x_k}} = \\
- \frac{1}{\rho} \frac{\partial \coh{p}}{ \partial x_i} + \nu \frac{\partial^2 \coh{u}_i}{\partial x_k \partial x_k} + \coh{f}_i.
\end{gathered}
\end{equation}
This leads to the budget equation for $\coh{u}_i$, i.e.
\begin{equation}
\begin{gathered}
\frac{\partial \coh{u}_i}{\partial t} + U_k \frac{\partial \coh{u}_i}{\partial x_k} + \coh{u}_k \frac{\partial U_i}{\partial x_k} + 
\frac{\partial}{\partial x_k} \left( \coh{u}_i \coh{u}_k - \aver{\coh{u}_i \coh{u}_k} \right) +
\frac{\partial}{\partial x_k} \left( \paver{u''_i u''_k} - \aver{u''_i u''_k} \right) =\\
-\frac{1}{\rho} \frac{\partial \coh{p}}{\partial x_i} + \nu \frac{\partial^2 \coh{u}_i}{\partial x_k \partial x_k} + \coh{f}_i.
\end{gathered}
\label{eq:CohUBudget}
\end{equation}

The budget equation for $u''_i$ is obtained by subtracting from \eqref{eq:Ns-3decomp} the budget equations for $U_i$ \eqref{eq:MeanUBudget} and $\coh{u}_i$ \eqref{eq:CohUBudget}:
\begin{equation}
\begin{gathered}
\frac{\partial u''_i}{\partial t} + U_k \frac{\partial u''_i}{\partial x_k} + \coh{u}_k \frac{\partial u''_i}{\partial x_k} + u''_k \frac{\partial U_i}{\partial x_k} + u''_k \frac{\partial \coh{u}_i}{\partial x_k} + \frac{\partial}{\partial x_k} \left( u_i'' u''_k - \paver{u''_i u''_k} \right) =\\
- \frac{1}{\rho} \frac{\partial p''}{\partial x_i} + \nu \frac{\partial^2 u''_i}{\partial x_k \partial x_k} + f''_i.
\end{gathered}
\label{eq:FluctUBudget}
\end{equation}

%----------------------------------------------------------------
\subsection{$\varphi$AGKE for $\delta \coh{u}_i \delta \coh{u}_j$}

The budget equation for $\coh{u}_i$ in $\vect{x}$ is subtracted from the one evaluated in $\vect{x}^+ =\vect{x} + \vect{r}$:
\begin{equation}
\begin{gathered}
\delta \left( \frac{\partial \coh{u}_i}{\partial t} \right) + \delta \left( U_k \frac{\partial \coh{u}_i}{\partial x_k} \right) +  
\delta \left( \coh{u}_k \frac{\partial U_i}{\partial x_k} \right) +
\delta \left( \frac{\partial}{\partial x_k} \left( \coh{u}_i \coh{u}_k - \aver{ \coh{u}_i \coh{u}_k } \right) \right) +  \\ +
\delta \left( \frac{\partial}{\partial x_k} \left( \paver{u''_i u''_k} - \aver{u''_i u''_k} \right) \right) =  
-\delta \left( \frac{1}{\rho} \frac{\partial \coh{p}}{\partial x_i} \right) + \delta \left( \nu \frac{\partial^2 \coh{u}_i}{ \partial x_k \partial x_k} \right) + \delta \left( \coh{f}_i \right).
\end{gathered}
\end{equation}
By recalling that the two reference systems are independent, one may write for example:
\begin{equation}
\delta \left( U_k \frac{\partial \coh{u}_i}{\partial x_k} \right) = U_k^+ \frac{\partial \delta \coh{u}_i}{\partial x_k^+} + U_k \frac{\partial \delta \coh{u}_i}{\partial x_k};
\end{equation}
using the same line of reasoning for all the other terms one obtains
\begin{equation}
\begin{gathered}
\frac{\partial \delta \coh{u}_i}{\partial t} + U_k^+ \frac{\partial \delta \coh{u}_i}{\partial x_k^+} + U_k \frac{\partial \delta \coh{u}_i}{\partial x_k} + \coh{u}_k^+ \frac{\partial \delta U_i}{\partial x_k^+} + \coh{u}_k \frac{\partial \delta U_i}{\partial x_k} + \coh{u}_k^+ \frac{\partial \delta \coh{u}_i}{\partial x_k^+} + \coh{u}_k \frac{\partial \delta \coh{u}_i}{\partial x_k} + \\
- \aver{\coh{u}_k^+ \frac{\partial \delta \coh{u}_i}{\partial x_k^+}} - \aver{\coh{u}_k \frac{\partial \delta \coh{u}_i}{\partial x_k}} +
\paver{ u_k^{''+} \frac{\partial \delta u''_i}{\partial x_k^+}} + \paver{u''_k \frac{\partial \delta u''_i}{\partial x_k}} - 
\aver{u_k^{''+} \frac{\partial \delta u''_i}{\partial x_k^+}} - \aver{u''_k \frac{\partial \delta u''_i}{\partial x_k}} = \\
- \frac{1}{\rho} \frac{\partial \delta \coh{p}}{\partial x_i^+}  - \frac{1}{\rho} \frac{\partial \delta \coh{p}}{\partial x_i} + 
\nu \left( \frac{\partial^2}{\partial x_k^+ \partial x_k^+} + \frac{\partial^2}{\partial x_k \partial x_k} \right) \delta \coh{u}_i + \delta \coh{f}_i.
\end{gathered}
\end{equation}

Then one may write for example
\begin{equation}
\coh{u}_k ^+ \frac{\partial \delta \coh{u}_i}{\partial x_k^+} = \delta \coh{u}_k \frac{\partial \delta \coh{u}_i}{\partial x_k^+} + \coh{u}_k \frac{\partial \delta \coh{u}_i}{\partial x_k^+}
\end{equation}
and using this expression for all the terms we obtain the budget equation for $\delta \coh{u}_i$:
\begin{equation}
\begin{gathered}
\frac{\partial \delta \coh{u}_i}{\partial t} +
\delta U_k \frac{\partial \delta \coh{u}_i}{\partial x_k^+} +  U_k \left( \frac{\partial}{\partial x_k^+} + \frac{\partial}{\partial x_k} \right) \delta \coh{u}_i + 
\delta \coh{u}_k \frac{\partial \delta U_i}{\partial x_k^+} + \coh{u}_k \left( \frac{\partial}{\partial x_k^+} + \frac{\partial}{\partial x_k} \right) \delta U_i + \\ +
\delta \coh{u}_k \frac{\partial \delta \coh{u}_i}{\partial x_k^+} + \coh{u}_k \left( \frac{\partial}{\partial x_k^+} + \frac{\partial}{\partial x_k} \right) \delta \coh{u}_i - 
\aver{\delta \coh{u}_k \frac{\partial \delta \coh{u}_i}{\partial x_k^+}} - \aver{\coh{u}_k \left( \frac{\partial}{\partial x_k^+} + \frac{\partial}{\partial x_k} \right) \delta \coh{u}_i } + \\ +
\paver{ \delta u''_k \frac{\partial \delta u''_i}{\partial x_k^+} } + \paver{ u''_k \left( \frac{\partial}{\partial x_k^+} + \frac{\partial}{\partial x_k} \right) \delta u''_i } - 
\aver{\delta u''_k \frac{\partial \delta u''_i}{\partial x_k^+}} - \aver{u''_k \left( \frac{\partial}{\partial x_k^+} + \frac{\partial}{\partial x_k} \right) \delta u''_i } = \\
- \frac{1}{\rho} \frac{\partial \delta \coh{p}}{\partial x_i^+} - \frac{1}{\rho} \frac{\partial \delta \coh{p}}{\partial x_i}
 + \nu \left( \frac{\partial^2}{\partial x_k^+ \partial x_k^+}  + \frac{\partial^2}{\partial x_k \partial x_k} \right) \delta \coh{u}_i  + \delta \coh{f}_i.
\end{gathered}
\end{equation}
This equation is multiplied by $\delta \coh{u}_j$ to obtain:
\begin{equation}
\begin{gathered}
\delta \coh{u}_j \frac{\partial \delta \coh{u}_i}{\partial t} +
\delta \coh{u}_j \delta U_k \frac{\partial \delta \coh{u}_i}{\partial x_k^+} +
\delta \coh{u}_j U_k \left( \frac{\partial}{\partial x_k^+} + \frac{\partial}{\partial x_k} \right) \delta \coh{u}_i + 
\delta \coh{u}_j \delta \coh{u}_k \frac{\partial \delta U_i}{\partial x_k^+} +  \\ +
\delta \coh{u}_j \coh{u}_k \left( \frac{\partial}{\partial x_k^+} + \frac{\partial}{\partial x_k} \right) \delta U_i + 
\delta \coh{u}_j \delta \coh{u}_k \frac{\partial \delta \coh{u}_i}{\partial x_k^+} + 
\delta \coh{u}_j \coh{u}_k \left( \frac{\partial}{\partial x_k^+} + \frac{\partial}{\partial x_k} \right) \delta \coh{u}_i - 
\delta \coh{u}_j \aver{\delta \coh{u}_k \frac{\partial \delta \coh{u}_i}{\partial x_k^+}} + \\ -
\delta \coh{u}_j \aver{\coh{u}_k \left( \frac{\partial}{\partial x_k^+} + \frac{\partial}{\partial x_k} \right) \delta \coh{u}_i } + 
\delta \coh{u}_j \paver{ \delta u''_k \frac{\partial \delta u''_i}{\partial x_k^+} } + 
\delta \coh{u}_j \paver{ u''_k \left( \frac{\partial}{\partial x_k^+} + \frac{\partial}{\partial x_k} \right) \delta u''_i } +\\
- \delta \coh{u}_j \aver{\delta u''_k \frac{\partial \delta u''_i}{\partial x_k^+}} - \delta \coh{u}_j \aver{u''_k \left( \frac{\partial}{\partial x_k^+} + \frac{\partial}{\partial x_k} \right) \delta u''_i } = 
- \delta \coh{u}_j \frac{1}{\rho} \left( \frac{\partial }{\partial x_i^+} + \frac{\partial }{\partial x_i} \right) \delta \coh{p} + \\ + 
\nu \delta \coh{u}_j \left( \frac{\partial^2}{\partial x_k^+ \partial x_k^+}  + \frac{\partial^2}{\partial x_k \partial x_k} \right) \delta \coh{u}_i + \delta \coh{u}_j \delta \coh{f}_i.
\end{gathered}
\end{equation}
The same equation is written again by swapping the $i$ and $j$ indices, and the two equations are then summed together:
\begin{equation}
\begin{gathered}
\frac{\partial}{\partial t} \delta \coh{u}_i \delta \coh{u}_j +
\delta \coh{u}_j \delta U_k \frac{\partial \delta \coh{u}_i}{\partial x_k^+} + \delta \coh{u}_i \delta U_k \frac{\partial \delta \coh{u}_j}{\partial x_k^+}  +
\delta \coh{u}_j U_k \left( \frac{\partial}{\partial x_k^+} + \frac{\partial}{\partial x_k} \right) \delta \coh{u}_i + \\ +
\delta \coh{u}_i U_k \left( \frac{\partial}{\partial x_k^+} + \frac{\partial}{\partial x_k} \right) \delta \coh{u}_j + 
\delta \coh{u}_j \delta \coh{u}_k \frac{\partial \delta U_i}{\partial x_k^+} + \delta \coh{u}_i \delta \coh{u}_k \frac{\partial \delta U_j}{\partial x_k^+}  +
\delta \coh{u}_j \coh{u}_k \left( \frac{\partial}{\partial x_k^+} + \frac{\partial}{\partial x_k} \right) \delta U_i + \\ +
\delta \coh{u}_i \coh{u}_k \left( \frac{\partial}{\partial x_k^+} + \frac{\partial}{\partial x_k} \right) \delta U_j + 
\delta \coh{u}_j \delta \coh{u}_k \frac{\partial \delta \coh{u}_i}{\partial x_k^+} + \delta \coh{u}_i \delta \coh{u}_k \frac{\partial \delta \coh{u}_j}{\partial x_k^+} + 
\delta \coh{u}_j \coh{u}_k \left( \frac{\partial}{\partial x_k^+} + \frac{\partial}{\partial x_k} \right) \delta \coh{u}_i + \\ +
\delta \coh{u}_i \coh{u}_k \left( \frac{\partial}{\partial x_k^+} + \frac{\partial}{\partial x_k} \right) \delta \coh{u}_j - 
\delta \coh{u}_j \aver{\delta \coh{u}_k \frac{\partial \delta \coh{u}_i}{\partial x_k^+}} - \delta \coh{u}_i \aver{\delta \coh{u}_k \frac{\partial \delta \coh{u}_j}{\partial x_k^+}} + \\ - 
\delta \coh{u}_j \aver{\coh{u}_k \left( \frac{\partial}{\partial x_k^+} + \frac{\partial}{\partial x_k} \right) \delta \coh{u}_i }  -
\delta \coh{u}_i \aver{\coh{u}_k \left( \frac{\partial}{\partial x_k^+} + \frac{\partial}{\partial x_k} \right) \delta \coh{u}_j } + \\ +
\delta \coh{u}_j \paver{ \delta u''_k \frac{\partial \delta u''_i}{\partial x_k^+} } + \delta \coh{u}_i \paver{ \delta u''_k \frac{\partial \delta u''_j}{\partial x_k^+} }  +
\delta \coh{u}_j \paver{ u''_k \left( \frac{\partial}{\partial x_k^+} + \frac{\partial}{\partial x_k} \right) \delta u''_i } + \\ +
\delta \coh{u}_i \paver{ u''_k \left( \frac{\partial}{\partial x_k^+} + \frac{\partial}{\partial x_k} \right) \delta u''_j } -
\delta \coh{u}_j \aver{\delta u''_k \frac{\partial \delta u''_i}{\partial x_k^+}} - \delta \coh{u}_j \aver{u''_k \left( \frac{\partial}{\partial x_k^+} + \frac{\partial}{\partial x_k} \right) \delta u''_i } + \\ - \delta \coh{u}_i \aver{\delta u''_k \frac{\partial \delta u''_j}{\partial x_k^+}} - \delta \coh{u}_i \aver{u''_k \left( \frac{\partial}{\partial x_k^+} + \frac{\partial}{\partial x_k} \right) \delta u''_j } = 
- \delta \coh{u}_j \frac{1}{\rho} \left( \frac{\partial }{\partial x_i^+} + \frac{\partial }{\partial x_i} \right) \delta \coh{p} + \\ 
- \delta \coh{u}_i \frac{1}{\rho} \left( \frac{\partial }{\partial x_j^+} + \frac{\partial }{\partial x_j} \right) \delta \coh{p} +  
\nu \delta \coh{u}_j \left( \frac{\partial^2}{\partial x_k^+ \partial x_k^+}  + \frac{\partial^2}{\partial x_k \partial x_k} \right) \delta \coh{u}_i + \\ +
\nu \delta \coh{u}_i \left( \frac{\partial^2}{\partial x_k^+ \partial x_k^+}  + \frac{\partial^2}{\partial x_k \partial x_k} \right) \delta \coh{u}_j + \delta \coh{u}_j \delta \coh{f}_i + \delta \coh{u}_i \delta \coh{f}_j.
\end{gathered}
\end{equation}

At this point, after applying the phase average operator $\paver{\cdot}$ and manipulating the equations, one obtains:
\begin{equation}
\begin{gathered}
\frac{\partial}{\partial t} \delta \coh{u}_i \delta \coh{u}_j +
\frac{\partial}{\partial x_k^+} \delta U_k \delta \coh{u}_i \delta \coh{u}_j  + 
\left( \frac{\partial}{\partial x_k^+} + \frac{\partial}{\partial x_k} \right) U_k \delta \coh{u}_i \delta \coh{u}_j + 
\delta \coh{u}_j \delta \coh{u}_k \frac{\partial \delta U_i}{\partial x_k^+} + 
\delta \coh{u}_i \delta \coh{u}_k  \frac{ \partial \delta U_j }{ \partial x_k^+ } + \\ +
\delta \coh{u}_j \coh{u}_k \left( \frac{\partial}{\partial x_k^+} + \frac{\partial}{\partial x_k} \right) \delta U_i + 
\delta \coh{u}_i \coh{u}_k \left( \frac{\partial}{\partial x_k^+} + \frac{\partial}{\partial x_k} \right) \delta U_j + 
\frac{\partial}{\partial x_k^+} \delta \coh{u}_k \delta \coh{u}_i \delta \coh{u}_j + \\ +
\left( \frac{\partial}{\partial x_k^+} + \frac{\partial}{\partial x_k} \right) \coh{u}_k \delta \coh{u}_i \delta \coh{u}_j   
- \delta \coh{u}_j \aver{\delta \coh{u}_k \frac{ \partial \delta \coh{u}_i}{\partial x_k^+} } -
  \delta \coh{u}_i \aver{\delta \coh{u}_k \frac{ \partial \delta \coh{u}_j}{\partial x_k^+} } +\\ -
  \delta \coh{u}_j \aver{ \coh{u}_k \left( \frac{\partial}{\partial x_k^+} + \frac{\partial}{\partial x_k} \right) \delta \coh{u}_i } - 
  \delta \coh{u}_i \aver{ \coh{u}_k \left( \frac{\partial}{\partial x_k^+} + \frac{\partial}{\partial x_k} \right) \delta \coh{u}_j } +  \\ + \delta \coh{u}_j \frac{\partial}{\partial x_k^+} \paver{\delta u''_i \delta u''_k}  + 
   \delta \coh{u}_i \frac{\partial}{\partial x_k^+} \paver{\delta u''_j \delta u''_k}  + 
 \delta \coh{u}_j \paver{ u''_k \left( \frac{\partial}{\partial x_k^+} + \frac{\partial}{\partial x_k} \right) \delta u''_i }  +  \\ +
  \delta \coh{u}_i \paver{ u''_k \left( \frac{\partial}{\partial x_k^+} + \frac{\partial}{\partial x_k} \right) \delta u''_j }
 -\delta \coh{u}_j \aver{\delta u''_k \frac{\partial \delta u''_i}{\partial x_k^+}} - \delta \coh{u}_j \aver{u''_k \left( \frac{\partial}{\partial x_k^+} + \frac{\partial}{\partial x_k} \right) \delta u''_i }+\\-\delta \coh{u}_i \aver{\delta u''_k \frac{\partial \delta u''_j}{\partial x_k^+}}
 - \delta \coh{u}_i \aver{u''_k \left( \frac{\partial}{\partial x_k^+} + \frac{\partial}{\partial x_k} \right) \delta u''_j }  = - \delta \coh{u}_j \frac{1}{\rho} \left( \frac{\partial}{\partial x_i^+} + \frac{\partial}{\partial x_i} \right) \delta \coh{p}  +\\ -
 \delta \coh{u}_i \frac{1}{\rho} \left( \frac{\partial}{\partial x_j^+} + \frac{\partial}{\partial x_j} \right) \delta \coh{p}  + 
 \nu  \delta \coh{u}_j \left( \frac{\partial^2}{\partial x_k^+ \partial x_k^+} + \frac{\partial^2}{\partial x_k \partial x_k} \right) \delta \coh{u}_i  + \\ +
 \nu \delta \coh{u}_i \left( \frac{\partial^2}{\partial x_k^+ \partial x_k^+} + \frac{\partial^2}{\partial x_k \partial x_k} \right) \delta \coh{u}_j + \delta \coh{u}_j \delta \coh{f}_i + \delta \coh{u}_i \delta \coh{f}_j .
\end{gathered}
\label{eq:a16}
\end{equation}

We now introduce the new independent variables $\vect{X}$ and $\vect{r}$ such that
\begin{equation*}
X_i = \frac{x_i+x_i^+}{2} \ \ \ \ \ r_i = x_i^+ - x_i.
\end{equation*}
As a result the $x_i$- and $x_i^+$-derivatives are related to the $X_i$- and $r_i$-derivatives by the following relations:
\begin{equation*}
\frac{\partial}{\partial x_i} = \frac{1}{2} \frac{\partial}{\partial X_i} - \frac{\partial}{\partial r_i}; \ \ \ \ 
\frac{\partial}{\partial x_i^+} = \frac{1}{2} \frac{\partial}{\partial X_i} + \frac{\partial}{\partial r_i}; \ \ \ \
\frac{\partial^2}{\partial x_k^+ \partial x_k^+} + \frac{\partial^2}{\partial x_k \partial x_k} = \frac{1}{2} \frac{\partial^2}{\partial X_k \partial X_k} + 2 \frac{\partial^2}{\partial r_k \partial r_k}
\end{equation*}
The previous equation \eqref{eq:a16} becomes:
\begin{equation}
\begin{gathered}
\frac{\partial}{\partial t} \delta \coh{u}_i \coh{u}_j + \left( \frac{1}{2} \frac{\partial}{\partial X_k} + \frac{\partial}{\partial r_k} \right) \delta U_k \delta \coh{u}_i \delta \coh{u}_j + 
\frac{ \partial }{ \partial X_k }  U_k \delta \coh{u}_i \delta \coh{u}_j + 
 \delta \coh{u}_j \delta \coh{u}_k  \left( \frac{1}{2} \frac{\partial}{\partial X_k} + \frac{\partial}{\partial r_k} \right) \delta U_i +  \\ +
 \delta \coh{u}_i \delta \coh{u}_k  \left( \frac{1}{2} \frac{\partial}{\partial X_k} + \frac{\partial}{\partial r_k} \right) \delta U_j + 
 \delta \coh{u}_j \coh{u}_k  \frac{\partial}{\partial X_k} \delta U_i + 
 \delta \coh{u}_i \coh{u}_k  \frac{\partial}{\partial X_k} \delta U_j + \\ +
\left( \frac{1}{2} \frac{\partial}{\partial X_k} + \frac{\partial}{\partial r_k} \right) ( \delta \coh{u}_k \delta \coh{u}_i \delta \coh{u}_j )  +
\frac{ \partial }{ \partial X_k} ( \coh{u}_k \delta \coh{u}_i \delta \coh{u}_j ) 
-\frac{\partial}{\partial r_k} \aver{\delta \coh{u}_i \delta \coh{u}_k} \delta \coh{u}_j  +
 \aver{\delta \coh{u}_i \delta \coh{u}_k } \frac{\partial \delta \coh{u}_j}{\partial r_k} + \\
-\frac{\partial }{ \partial X_k} \aver{ \coh{u}_k^* \delta \coh{u}_i } \delta \coh{u}_j  +
 \aver{\coh{u}_k^* \delta \coh{u}_i } \frac{\partial \delta \coh{u}_j}{\partial X_k} 
-\frac{\partial }{\partial r_k} \aver{\delta \coh{u}_j \delta \coh{u}_k}  \delta \coh{u}_i +
 \aver{\delta \coh{u}_j \delta \coh{u}_k } \frac{\partial \delta \coh{u}_i}{\partial r_k} 
-\frac{\partial }{ \partial X_k} \aver{\coh{u}_k^* \delta \coh{u}_j} \delta \coh{u}_i + \\ +
 \aver{\coh{u}_k^* \delta \coh{u}_j } \frac{\partial \delta \coh{u}_i}{\partial X_k} +
\delta \coh{u}_j \left( \frac{1}{2} \frac{\partial}{\partial X_k} + \frac{\partial}{\partial r_k} \right) \paver{\delta u''_i \delta u''_k}  + 
 \delta \coh{u}_i \left( \frac{1}{2} \frac{\partial}{\partial X_k} + \frac{\partial}{\partial r_k} \right) \paver{\delta u''_j \delta u''_k}   + \\ +
 \delta \coh{u}_j \paver{ u''_k \frac{\partial}{\partial X_k} \delta u''_i }  +
 \delta \coh{u}_i \paver{ u''_k \frac{\partial}{\partial X_k} \delta u''_j }   
 -\frac{\partial}{\partial r_k} \aver{\delta u''_i \delta u''_k} \delta \coh{u}_j  +
 \aver{\delta u''_i \delta u''_k } \frac{\partial \delta \coh{u}_j}{\partial r_k} + \\
-\frac{\partial }{ \partial X_k} \aver{ u_k^{''*} \delta u''_i } \delta \coh{u}_j  +
 \aver{u''^*_k \delta u''_i } \frac{\partial \delta \coh{u}_j}{\partial X_k} 
-\frac{\partial }{\partial r_k} \aver{\delta u''_j \delta u''_k}  \delta \coh{u}_i + \\ +
 \aver{\delta u''_j \delta u''_k } \frac{\partial \delta \coh{u}_i}{\partial r_k} 
-\frac{\partial }{ \partial X_k} \aver{u''^*_k \delta u''_j} \delta \coh{u}_i  +
 \aver{u''^*_k \delta u''_j } \frac{\partial \delta \coh{u}_i}{\partial X_k} = \\
- \delta \coh{u}_j \frac{1}{\rho} \left( \frac{\partial}{\partial X_i} \right) \delta \coh{p}  - \delta \coh{u}_i \frac{1}{\rho} \left( \frac{\partial}{\partial X_j} \right) \delta \coh{p}  +
\nu  \delta \coh{u}_j \left( \frac{1}{2} \frac{\partial^2}{\partial X_k \partial X_k} + 2 \frac{\partial^2}{\partial r_k \partial r_k} \right) \delta \coh{u}_i  + \\ +
\nu  \delta \coh{u}_i \left( \frac{1}{2} \frac{\partial^2}{\partial X_k \partial X_k} + 2 \frac{\partial^2}{\partial r_k \partial r_k} \right) \delta \coh{u}_j + \delta \coh{u}_j \delta \coh{f}_i + \delta \coh{u}_i \delta \coh{f}_j
\end{gathered}
\end{equation}
where the star $(\cdot)^*$ denotes the average of any quantity between $\vect{x}$ and $\vect{x}^+$. We also observe that:
\begin{equation}
\begin{gathered}
 \delta \coh{u}_j \left( \frac{1}{2} \frac{\partial}{\partial X_k} + \frac{\partial}{\partial r_k} \right) \paver{ \delta u''_i \delta u''_k} +
 \delta \coh{u}_j \paver{ u''_k \frac{\partial}{\partial X_k} \delta u''_i}  = \\
\delta \coh{u}_j \left( \frac{1}{2} \frac{\partial}{\partial X_k} + \frac{\partial}{\partial r_k} \right) \paver{ \delta u''_i \delta u''_k}  +
 \delta \coh{u}_j \frac{ \partial }{\partial X_k} \paver{u''_k \delta u''_i}  = \\
\delta \coh{u}_j \frac{\partial}{\partial r_k} \paver{\delta u''_i \delta u''_k}  +
\delta \coh{u}_j \frac{\partial}{\partial X_k} \paver{\frac{1}{2} \left( u''_k + u_k^{''+} \right) \delta u''_i } = \\
\frac{\partial}{\partial r_k}  \paver{\delta u''_i \delta u''_k} \delta \coh{u}_j  - \paver{\delta u''_i \delta u''_k} \frac{\partial \delta \coh{u}_j}{\partial r_k}  + 
\frac{\partial}{\partial X_k} \paver{ u_k^{\prime \prime *} \delta u''_i } \delta \coh{u}_j  - \paver{ u_k^{\prime \prime *} \delta u''_i} \frac{\partial}{\partial X_k} \delta \coh{u}_j   
\end{gathered}
\end{equation}

The viscous term can be simplified as:
\begin{equation}
\begin{gathered}
\nu \delta \coh{u}_j \left( \frac{1}{2} \frac{\partial^2}{\partial X_k \partial X_k} + 2 \frac{\partial^2}{\partial r_k \partial r_k} \right) \delta \coh{u}_i  + 
\nu \delta \coh{u}_i \left( \frac{1}{2} \frac{\partial^2}{\partial X_k \partial X_k} + 2 \frac{\partial^2}{\partial r_k \partial r_k} \right) \delta \coh{u}_j  = \\
\frac{\nu}{2} \frac{\partial^2}{\partial X_k \partial X_k}\delta \coh{u}_i \delta \coh{u}_j  + 2 \nu \frac{\partial^2}{\partial r_k \partial r_k} \delta \coh{u}_i \delta \coh{u}_j -
\nu  \frac{\partial \delta \coh{u}_i}{\partial X_k} \frac{\partial \delta \coh{u}_j}{\partial X_k}  - 
4 \nu  \frac{\partial \delta \coh{u}_i}{\partial r_k} \frac{\partial \delta \coh{u}_j}{\partial r_k}  = \\
\frac{\nu}{2} \frac{\partial^2}{\partial X_k \partial X_k}  \delta \coh{u}_i \delta \coh{u}_j  +
2 \nu \frac{\partial^2}{\partial r_k \partial r_k}  \delta \coh{u}_i \delta \coh{u}_j  - 2 \left( \epsilon^{c+}_{ij} + \epsilon^c_{ij} \right)
\end{gathered}
\end{equation}
where $\epsilon^{c}_{ij}$ is the pseudo-dissipation tensor of the coherent part of the velocity, defined as:
\begin{equation}
\epsilon^{c}_{ij} = \nu \aver{ \frac{\partial \coh{u}_i}{\partial x_k} \frac{\partial \coh{u}_j} {\partial x_k} }
\end{equation}
Moreover we write:
\begin{equation}
\begin{gathered}
\delta \coh{u}_j \delta \coh{u}_k \frac{\partial \delta U_i}{\partial r_k} =  \delta \coh u_j \delta \coh u_k \left( \frac{\partial U_i}{\partial x_k} \right)^*
\end{gathered}
\end{equation}
and:
\begin{equation}
  \delta \coh{u}_j \coh{u}_k^* \frac{\partial}{\partial X_k} \delta U_i = \delta \coh{u}_j \coh{u}_k^* \delta \left ( \frac{\partial U_i}{\partial x_k} \right) .
\end{equation}

Finally, the budget equation for $ \delta \coh{u}_i \delta \coh{u}_j $ is obtained:
\begin{equation}
\begin{gathered}
\frac{\partial}{\partial t} \delta \coh{u}_i \delta \coh{u}_j + 
\frac{\partial}{\partial r_k}  \delta U_k \delta \coh{u}_i \delta \coh{u}_j  +
\frac{\partial}{\partial X_k}  U_k^*      \delta \coh{u}_i \delta \coh{u}_j  +
\frac{\partial}{\partial r_k}  \delta \coh{u}_k \delta \coh{u}_i \delta \coh{u}_j  + 
\frac{\partial}{\partial X_k}  \coh{u}_k^*      \delta \coh{u}_i \delta \coh{u}_j  + \\ +
\frac{\partial}{\partial r_k}  \paver{\delta u''_k \delta u''_i} \delta \coh{u}_j  + 
\frac{\partial}{\partial X_k}  \paver{ u''^*_k     \delta u''_i} \delta \coh{u}_j  + 
\frac{\partial}{\partial r_k}  \paver{\delta u''_k \delta u''_j} \delta \coh{u}_i  + 
\frac{\partial}{\partial X_k}  \paver{ u''^*_k     \delta u''_j} \delta \coh{u}_i  + \\ -
2 \nu \frac{\partial^2}{\partial r_k \partial r_k} \delta \coh{u}_i \delta \coh{u}_j  - 
\frac{\nu}{2} \frac{\partial}{\partial X_k} \left( \frac{\partial}{\partial X_k} \delta \coh{u}_i \delta \coh{u}_j \right) + 
\frac{\partial}{\partial X_i} \frac{1}{\rho} \delta \coh{p} \delta \coh{u}_j + 
\frac{\partial}{\partial X_j} \frac{1}{\rho} \delta \coh{p} \delta \coh{u}_i + \\ 
-\frac{\partial}{\partial r_k} \aver{\delta \coh{u}_i \delta \coh{u}_k} \delta \coh{u}_j -
 \frac{\partial}{\partial X_k}  \aver{ \coh{u}_k^* \delta \coh{u}_i} \delta \coh{u}_j -
 \frac{\partial}{\partial r_k} \aver{\delta \coh{u}_j \delta \coh{u}_k} \delta \coh{u}_i -
 \frac{\partial}{\partial X_k} \aver{\coh{u}_k^* \delta \coh{u}_j} \delta \coh{u}_i + \\ 
-\frac{\partial}{\partial r_k} \aver{\delta u''_i \delta u''_k} \delta \coh{u}_j -
 \frac{\partial}{\partial X_k}  \aver{ u''^*_k \delta u''_i} \delta \coh{u}_j -
 \frac{\partial}{\partial r_k} \aver{\delta u''_j \delta u''_k} \delta \coh{u}_i -
 \frac{\partial}{\partial X_k} \aver{u''^*_k \delta u''_j} \delta \coh{u}_i= \\
- \delta \coh{u}_j \delta \coh{u}_k \left( \frac{\partial U_i}{\partial x_k} \right)^* -
  \delta \coh{u}_i \delta \coh{u}_k \left( \frac{\partial U_j}{\partial x_k} \right)^* -
  \delta \coh{u}_j \coh{u}_k^* \delta \left( \frac{\partial U_i}{\partial x_k} \right) -
  \delta \coh{u}_i \coh{u}_k^* \delta \left( \frac{\partial U_j}{\partial x_k} \right) + \\ 
  - \aver{\delta \coh{u}_i \delta \coh{u}_k } \left( \frac{\partial \coh{u}_j}{\partial x_k} \right)^* -
  \aver{\delta \coh{u}_j \delta \coh{u}_k } \left( \frac{\partial \coh{u}_i}{\partial x_k} \right)^* -
  \aver{\delta \coh{u}_i \coh{u}_k^* } \delta \left( \frac{\partial \coh{u}_j}{\partial x_k} \right) -
  \aver{\delta \coh{u}_j \coh{u}_k^* } \delta \left( \frac{\partial \coh{u}_i}{\partial x_k} \right) + \\ 
- \aver{\delta u''_i \delta u''_k } \left( \frac{\partial \coh{u}_j}{\partial x_k} \right)^* -
  \aver{\delta u''_j \delta u''_k } \left( \frac{\partial \coh{u}_i}{\partial x_k} \right)^* -
  \aver{\delta u''_i u''^*_k } \delta \left( \frac{\partial \coh{u}_j}{\partial x_k} \right) -
  \aver{\delta u''_j u''^*_k } \delta \left( \frac{\partial \coh{u}_i}{\partial x_k} \right) + \\ +
  \paver{ \delta u''_i \delta u''_k}  \frac{\partial \delta \coh{u}_j}{\partial r_k}  +
  \paver{ \delta u''_i u''^*_k }      \frac{\partial \delta \coh{u}_j}{\partial X_k}  + 
  \paver{ \delta u''_j \delta u''_k} \frac{\partial \delta \coh{u}_i}{\partial r_k}  +
  \paver{ \delta u''_j u''^*_k    }  \frac{\partial \delta \coh{u}_i}{\partial X_k}  + \\ +
  \frac{1}{\rho} \delta \coh{p} \frac{\partial \delta \coh{u}_j}{\partial X_i}  +
  \frac{1}{\rho} \delta \coh{p} \frac{\partial \delta \coh{u}_i}{\partial X_j}  - 
  4 \epsilon_{ij}^{c*} + \delta \coh{u}_j \delta \coh{f}_i + \delta \coh{u}_i \delta \coh{f}_j.
\end{gathered}
\end{equation}

%-----------------------------------------------------------------
\subsection{$\varphi$AGKE for $\paver{\delta u''_i \delta u''_j}$}

We write the budget equation for $u''_i$ twice for the positions $\vect{x}$ and $\vect{x}^+=\vect{x}+\vect{r}$, then the first is subtracted from the second:
\begin{equation}
\begin{gathered}
  \delta \left( \frac{\partial u''_i}{\partial t} \right) + \delta \left( U_k \frac{\partial u''_i}{\partial x_k} \right) +
  \delta \left( \coh{u}_k \frac{\partial u''_i}{\partial x_k} \right)+
  \delta \left( u''_k \frac{\partial U_i}{\partial x_k} \right) + \delta \left( u''_k \frac{\partial \coh{u}_i}{\partial x_k} \right) + \\ +
  \delta \left( \frac{\partial}{\partial x_k} \left( u''_i u''_k - \paver{u''_i u''_k} \right) \right) = 
  - \delta \left( \frac{1}{\rho} \frac{\partial p''}{\partial x_i} \right) + \delta \left( \nu \frac{\partial^2 u''_i}{\partial x_k \partial x_k} \right) + \delta \left( f''_i\right).
\end{gathered}
\end{equation}
Following the line of reasoning described above, the equation for $\delta u''_i$ is obtained, i.e:
\begin{equation}
\begin{gathered}
\frac{\partial \delta u''_i}{\partial t} + \delta U_k \frac{\partial \delta u''_i}{\partial x_k^+} + 
U_k \frac{\partial \delta u''_i}{\partial x_k^+} + U_k \frac{\partial \delta u''_i}{\partial x_k} + 
\delta \coh{u}_k \frac{\partial \delta u''_i}{\partial x_k^+} + \coh{u}_k \frac{\partial \delta u''_i}{\partial x_k^+} + \coh{u}_k \frac{\partial \delta u''_i}{\partial x_k} +  \\ +
\delta u''_k \frac{\partial \delta U_i}{\partial x_k^+} + u''_k \frac{\partial \delta U_i}{\partial x_k^+} + u''_k \frac{\partial \delta U_i}{\partial x_k} + 
\delta u''_k \frac{\partial \delta \coh{u}_i}{\partial x_k^+} + u''_k \frac{\partial \delta \coh{u}_i}{\partial x_k^+} + u''_k \frac{\partial \delta \coh{u}_i}{\partial x_k} +  
\delta u''_k \frac{\partial \delta u''_i}{\partial x_k^+} + \\ +
 u''_k \frac{\partial \delta u''_i}{\partial x_k^+} + u''_k \frac{\partial \delta u''_i}{\partial x_k} - 
\paver{\delta u''_k \frac{\partial \delta u''_i}{\partial x_k^+} } - \paver{ u''_k \left( \frac{\partial}{\partial x_k^+} + \frac{\partial}{\partial x_k} \right) \delta u''_i } = \\
- \frac{1}{\rho} \frac{\partial \delta p''}{\partial x_i^+} - \frac{1}{\rho} \frac{\partial \delta p''}{\partial x_i} + 
\nu \left( \frac{\partial^2}{\partial x_k^+ \partial x_k^+} + \frac{\partial^2}{\partial x_k \partial x_k} \right) \delta u_i +\delta f''_i.
\end{gathered}
\end{equation}
As above, we first multiply this equation for $\delta u''_j$, and then we sum to the  same equation with swapped $i$ and $j$ indices. Using again the independence of the $\vect{x}$ and $\vect{x}^+$ reference systems and incompressibility, and applying the phase average operator $\paver{\cdot}$ we obtain:
\begin{equation}
\begin{gathered}
\frac{\partial}{\partial t} \paver{\delta u''_i \delta u''_j } +
\frac{\partial}{\partial x_k^+} \delta U_k\paver{ \delta u''_i \delta u''_j} + 
\left( \frac{\partial}{\partial x_k^+} + \frac{\partial}{\partial x_k} \right) U_k\paver{ \delta u''_i \delta u''_j} +
\paver{\delta u''_j \delta u''_k} \frac{\partial \delta U_i}{\partial x_k^+} + \paver{\delta u''_i \delta u''_k} \frac{\partial \delta U_j}{\partial x_k^+} + \\ +
\paver{\delta u''_j u''_k} \left( \frac{\partial}{\partial x_k^+} + \frac{\partial}{\partial x_k} \right) \delta U_i +
\paver{\delta u''_i u''_k} \left( \frac{\partial}{\partial x_k^+} + \frac{\partial}{\partial x_k} \right) \delta U_j + 
\frac{\partial}{\partial x_k^+} \delta \coh{u}_k \paver{\delta u''_j  \delta u''_i} +  \\ +
\left( \frac{\partial}{\partial x_k^+} + \frac{\partial}{\partial x_k} \right) \coh{u}_k \paver{\delta u''_j  \delta u''_i} + 
 \paver{\delta u''_j \delta u''_k} \frac{\partial \delta \coh{u}_i}{\partial x_k^+}  +  
 \paver{\delta u''_i \delta u''_k} \frac{\partial \delta \coh{u}_j}{\partial x_k^+}  +  
 \paver{\delta u''_j u''_k} \left( \frac{\partial}{\partial x_k^+} + \frac{\partial}{\partial x_k} \right) \delta \coh{u}_i  + \\ +
 \paver{\delta u''_i u''_k} \left( \frac{\partial}{\partial x_k^+} + \frac{\partial}{\partial x_k} \right) \delta \coh{u}_j  + 
\frac{\partial}{\partial x_k^+} \paver{\delta u''_k \delta u''_i \delta u''_j} + 
\left( \frac{\partial}{\partial x_k^+} + \frac{\partial}{\partial x_k} \right) \paver{u''_k \delta u''_i \delta u''_j} = \\ 
- \frac{1}{\rho} \left( \frac{\partial}{\partial x_i^+} + \frac{\partial}{\partial x_i} \right) \paver{\delta p'' \delta u''_j} - 
  \frac{1}{\rho} \left( \frac{\partial}{\partial x_j^+} + \frac{\partial}{\partial x_j} \right) \paver{\delta p'' \delta u''_i} + \\ +
\paver{ \frac{1}{\rho} \delta p'' \left( \frac{\partial}{\partial x_i^+} + \frac{\partial}{\partial x_i} \right) \delta u''_j } + 
\paver{ \frac{1}{\rho} \delta p'' \left( \frac{\partial}{\partial x_j^+} + \frac{\partial}{\partial x_j} \right) \delta u''_i } + \\ +
\nu \paver{\delta u''_j \left( \frac{\partial^2}{\partial x_k^+ \partial x_k^+} + \frac{\partial^2}{\partial x_k \partial x_k} \right) \delta u''_i} + 
\nu \paver{\delta u''_i \left( \frac{\partial^2}{\partial x_k^+ \partial x_k^+} + \frac{\partial^2}{\partial x_k \partial x_k} \right) \delta u''_j}
+ \paver{\delta f''_i  \delta u''_j} + \paver{\delta f''_j  \delta u''_i}.
\end{gathered}
\end{equation}

We switch as above to the notation with $\vect{X}$ and $\vect{r}$ to obtain:
\begin{equation}
\begin{gathered}
\frac{\partial}{\partial t} \paver{\delta u''_i \delta u''_j} + 
\frac{\partial}{\partial r_k} U_k\paver{\delta  \delta u''_i \delta u''_j} + \frac{\partial}{\partial X_k} U_k^*\paver{ \delta u''_i \delta u''_j} +    
\paver{\delta u''_j \delta u''_k} \frac{\partial \delta U_i}{\partial r_k} + 
\paver{\delta u''_i \delta u''_k} \frac{\partial \delta U_j}{\partial r_k} + \\ +   
\paver{\delta u''_j u''^*_k} \frac{\partial \delta U_i}{\partial X_k} + 
\paver{\delta u''_i u''^*_k} \frac{\partial \delta U_j}{\partial X_k} + 
\paver{\delta u''_j \delta u''_k} \frac{\partial \delta \coh{u}_i}{\partial r_k} + 
\paver{\delta u''_j u''^*_k} \frac{\partial \delta \coh{u}_i}{\partial X_k} + 
\paver{\delta u''_i \delta u''_k} \frac{\partial \delta \coh{u}_j}{\partial r_k} + \\ +
\paver{\delta u''_j u''^*_k} \frac{\partial \delta \coh{u}_j}{\partial X_k} + 
\frac{\partial}{\partial r_k} \paver{\delta u''_k \delta u''_i \delta u''_j} +
\frac{\partial}{\partial X_k} \paver{u''^*_k \delta u''_i \delta u''_j} +
\frac{\partial}{\partial r_k} \delta \coh{u}_k \paver{ \delta u''_i \delta u''_j} + \\ +  
\frac{\partial}{\partial X_k} \coh{u}^*_k\paver{ \delta u''_i \delta u''_j} + 
\frac{\partial}{\partial X_i} \frac{1}{\rho} \paver{\delta p'' \delta u''_j} + 
\frac{\partial}{\partial X_j} \frac{1}{\rho} \paver{\delta p'' \delta u''_i} = 
\frac{1}{\rho}\paver{ \delta p'' \frac{\partial \delta u''_j}{\partial X_i}} + \\ +
\frac{1}{\rho}\paver{ \delta p'' \frac{\partial \delta u''_i}{\partial X_j}} + 
\frac{\nu}{2} \frac{\partial^2}{\partial X_k \partial X_k} \paver{\delta u''_i \delta u''_j} + 2 \nu \frac{\partial^2}{\partial r_k \partial r_k} \paver{\delta u''_i \delta u''_j} - 
2 \left( \epsilon_{ij}^{s+} + \epsilon^s_{ij} \right) + \paver{\delta f''_i  \delta u''_j} + \paver{\delta f''_j  \delta u''_i}
\end{gathered}
\end{equation}
where 
\begin{equation}
\epsilon^s_{ij}=\nu \paver{\frac{\partial u''_i}{\partial x_k}\frac{\partial u''_j}{\partial x_k}}.
\end{equation}
is the pseudo-dissipation tensor of the stochastic part of the velocity.
Also in this case we can write
\begin{equation}
\paver{\delta u''_j \delta u''_k} \frac{\partial \delta U_i}{\partial r_k} = \paver{\delta u''_j \delta u''_k} \left( \frac{\partial U_i}{\partial x_k} \right)^*
\end{equation}
and
\begin{equation}
\paver{\delta u''_j u''^*_k} \frac{\partial \delta U_i}{\partial X_k} = \paver{\delta u''_j u''^*_k} \delta \left( \frac{\partial U_i}{\partial x_k} \right)
\end{equation}
so that the budget equation for $\paver{\delta u''_i \delta u''_j}$ is eventually obtained:
\begin{equation}
\begin{gathered}
\frac{\partial}{\partial t} \paver{\delta u''_i \delta u''_j} + 
\frac{\partial}{\partial r_k} \delta U_k \paver{\delta u''_i \delta u''_j} +\frac{\partial}{\partial X_k} U_k^* \paver{\delta u''_i \delta u''_j} +
\frac{\partial}{\partial r_k} \paver{\delta u''_k \delta u''_i \delta u''_j} + \frac{\partial}{\partial X_k} \paver{u''^*_k \delta u''_i \delta u''_j} + \\ +
\frac{\partial}{\partial r_k} \left( -2 \nu \frac{\partial}{\partial r_k} \paver{\delta u''_i \delta u''_j} \right) + 
\frac{\partial}{\partial X_k} \left( -\frac{\nu}{2} \frac{\partial}{\partial X_k} \paver{\delta u''_i \delta u''_j} \right) + 
\frac{\partial}{\partial r_k} \delta \coh{u}_k \paver{\delta u''_i \delta u''_j} + 
\frac{\partial}{\partial X_k} \coh{u}_k^* \paver{\delta u''_i \delta u''_j} + \\ +
\frac{\partial}{\partial X_i} \frac{1}{\rho} \paver{\delta p'' \delta u''_j} + \frac{\partial}{\partial X_j} \frac{1}{\rho} \paver{\delta p'' \delta u''_i} = 
- \paver{\delta u''_j \delta u''_k} \left( \frac{\partial U_i}{\partial x_k} \right)^* - \paver{\delta u''_i \delta u''_k} \left( \frac{ \partial U_j}{\partial X_k} \right)^*  + \\ - 
\paver{\delta u''_j u''^*_k} \delta \left( \frac{\partial U_i}{\partial x_k} \right)  - \paver{\delta u''_i u''^*_k} \delta \left( \frac{\partial U_j}{\partial x_k} \right)  - 
 \paver{\delta u''_j \delta u''_k} \left( \frac{\partial \coh{u}_i}{\partial x_k} \right)^* - 
 \paver{\delta u''_i \delta u''_k} \left( \frac{\partial \coh{u}_j}{\partial x_k} \right)^* + \\ -
 \paver{\delta u''_j u''^*_k} \delta \left( \frac{\partial \coh{u}_i}{\partial x_k} \right)  -
 \paver{\delta u''_i u''^*_k} \delta \left( \frac{\partial \coh{u}_j}{\partial x_k} \right)  + 
\frac{1}{\rho}\paver{ \delta p'' \frac{\partial \delta u''_j}{\partial X_i} } +
\frac{1}{\rho}\paver{ \delta p'' \frac{\partial \delta u''_i}{\partial X_j} } - 4 \epsilon_{ij}^{s*}
+ \paver{\delta f''_i  \delta u''_j} + \paver{\delta f''_j  \delta u''_i}
\end{gathered}
\end{equation}

\section{The $\varphi$AGKE for the plane channel flow with oscillating walls}
\label{sec:tcf-phiagke}

The special form assumed by the $\varphi$AGKE under the symmetries of a plane channel flow with spanwise oscillations is reported below. The coherent part reduces to:

\begin{equation} 
\label{eq:dcohuidcohuj}
\begin{gathered}
\omega \frac{\partial \delta \coh{u}_i \delta \coh{u}_j}{\partial \varphi} +
\frac{\partial}{\partial r_k} \underbrace{ \left( \paver{\delta u''_k \delta u_i'' } \delta \coh{u}_j + \paver{\delta u''_k \delta u_j'' } \delta \coh{u}_i \right) }_{\text{Turbulent transport} } +
\frac{\partial}{\partial r_y} \underbrace{ \left(-2 \nu \frac{\partial  \delta \coh{u}_i \delta \coh{u}_j }{\partial r_y} \right)}_{\text{Viscous diffusion}} + 
 \frac{\partial}{\partial Y} \underbrace{ \left(- \frac{\nu}{2} \frac{\partial \delta \coh{u}_i \delta \coh{u}_j }{\partial Y} \right)}_{\text{Viscous diffusion}} + \\ +
\frac{\partial}{\partial Y} \underbrace{\left( \paver{ v^{\prime \prime *}   \delta u_i''} \delta \coh{u}_j + \paver{ v^{\prime \prime *}   \delta u_j''} \delta \coh{u}_i \right)}_{\text{Turbulent transport}}
+ \frac{\partial}{\partial Y}  \underbrace{ \left( \frac{1}{\rho} \delta \coh{p} \delta \coh{u}_i \delta_{j2} \right) }_{ \text{Pressure transport}}
+ \frac{\partial}{\partial Y}  \underbrace{ \left( \frac{1}{\rho} \delta \coh{p} \delta \coh{u}_j \delta_{i2} \right) }_{ \text{Pressure transport}} 
 = \\
-\underbrace{\left[ -\paver{\delta u_i''  \delta v''}        \left( \frac{\partial \coh{u}}{\partial y} \right)^* \delta_{j1} - \paver{\delta u_j''  \delta v''}        \left( \frac{\partial \coh{u}}{\partial y} \right)^* \delta_{i1} - \paver{\delta u_i''  v^{ \prime \prime *} }    \delta \left( \frac{\partial \coh{u}}{\partial y} \right) \delta_{j1} - \paver{\delta u_j''  v^{ \prime \prime *} }    \delta \left( \frac{\partial \coh{u}}{\partial y} \right) \delta_{i1} \right]}_{\cs{\pph}_{ij}} + \\
-\underbrace{\left[ -\paver{\delta u_i''  \delta v''}        \left( \frac{\partial \coh{w}}{\partial y} \right)^* \delta_{j3} - \paver{\delta u_j''  \delta v''}        \left( \frac{\partial \coh{w}}{\partial y} \right)^* \delta_{i3} - \paver{\delta u_i''  v^{ \prime \prime *} }    \delta \left( \frac{\partial \coh{w}}{\partial y} \right) \delta_{j3} - \paver{\delta u_j''  v^{ \prime \prime *} }    \delta \left( \frac{\partial \coh{w}}{\partial y} \right) \delta_{i3} \right]}_{\cs{\pph}_{ij}} + \\
+\underbrace{ \frac{1}{\rho} \delta \coh{p} \frac{ \partial \delta \coh{u}_i }{ \partial Y} \delta_{j2} 
            + \frac{1}{\rho} \delta \coh{p} \frac{ \partial \delta \coh{u}_j }{ \partial Y} \delta_{i2} }_{\pi^c_{ij}}
 \underbrace{- \ 4 \epsilon_{ij}^{c*} }_{\dph_{ij}^c} + \\
+ \underbrace{ \frac{\partial}{\partial r_k} \left[ \aver{ \delta u''_i \delta u''_k} \delta \coh{u}_j + \aver{\delta u''_j \delta u''_k} \delta \coh{u}_i \right] +
\frac{\partial}{\partial Y} \left[ \aver{v^{\prime \prime *} \delta u''_i} \delta \coh{u}_j + \aver{v^{\prime \prime *} \delta u''_j} \delta \coh{u}_i \right] }_{\zeta^c_{ij} }  + \\
+ \underbrace{ \left[ - \aver{\delta u''_i \delta v''}  \left( \frac{\partial \coh{u}_j}{\partial y} \right)^* 
                      - \aver{\delta u''_j \delta v''}  \left( \frac{\partial \coh{u}_i}{\partial y} \right)^*
                      - \aver{\delta u''_i v^{\prime \prime *}} \delta \left( \frac{\partial \coh{u}_j}{\partial y} \right) 
                      - \aver{\delta u''_j v^{\prime \prime *}} \delta \left( \frac{\partial \coh{u}_i}{\partial y} \right) \right]}_{\zeta^c_{ij}}.
\end{gathered}
\end{equation}

The $\varphi$AGKE for the stochastic part $\paver{\delta u''_i \delta u''_j}$ become:
\begin{equation} \label{eq:duiduj}
\begin{gathered}
\omega \frac{\partial \paver{\delta u''_i \delta u''_j} }{\partial \varphi} +
\frac{\partial}{\partial r_x}\underbrace{ \left( \delta U \paver{ \delta u_i'' \delta u_j'' } \right) }_{\text{Mean transport}} + \frac{\partial}{\partial r_x}\underbrace{ \left( \delta \coh{u} \paver{\delta u_i'' \delta u_j'' } \right) }_{\text{Coherent transport}} + 
\frac{\partial}{\partial r_z}\underbrace{ \left( \delta \coh{w} \paver{\delta u_i'' \delta u_j'' } \right) }_{\text{Coherent transport}} + \\ + 
\frac{\partial}{\partial r_k}\underbrace{ \left( \paver{ \delta u''_k \delta u_i''  \delta u_j'' } \right) }_{\text{Turbulent transport}}  +  
\frac{\partial}{\partial r_k}\underbrace{ \left( -2 \nu \frac{\partial \paver{ \delta u_i'' \delta u_j'' } }{\partial r_k} \right)}_{\text{Viscous diffusion}}  + 
\frac{\partial}{\partial Y}\underbrace{ \left( \paver{ v^{\prime \prime *}   \delta u_i'' \delta u_j'' } \right) }_{\text{Turbulent transport}} + \\ +
\frac{\partial}{\partial Y}\underbrace{ \left( - \frac{\nu}{2} \frac{\partial \paver{\delta u_i'' \delta u_j''} }{\partial Y} \right)}_{\text{Viscous diffusion}} +
\frac{\partial}{\partial Y}\underbrace{ \left( \frac{1}{\rho}  \paver{    \delta p'' \delta u_j'' } \delta_{i2} + \frac{1}{\rho} \paver{ \delta p'' \delta u_i''} \delta_{j2} \right) }_{\text{Pressure transport}} = \\ 
\underbrace{ \left[-  \paver{ \delta u_i'' \delta v''} \left( \frac{d U}{d y} \right)^* \delta_{j1}
            -  \paver{ \delta u_j'' \delta v'' } \left( \frac{d U}{d y} \right)^* \delta_{i1}  
- \paver{ \delta u_i'' v^{\prime \prime *} } \delta \left( \frac{d U}{d y} \right) \delta_{j1} 
- \paver{  \delta u_j'' v^{\prime \prime *} } \delta \left( \frac{d U}{d y} \right) \delta_{i1} \right] }_{\ms{\pph}_{ij}} + \\ 
%\underbrace{-2 \paver{\delta u'' \delta v''} \left( \frac{d \coh{u}}{d y} \right)^* \delta_{i1} \delta_{j1}
%-2 \paver{\delta u'' v^{\prime \prime *}  }   \delta \left( \frac{d  \coh{u}}{d y} \right) \delta_{i1} \delta_{j1}}_{\cs{\pph}_{11}}  + \\ 
%\underbrace{-2 \paver{\delta v'' \delta w''} \left( \frac{d \coh{w}}{d y} \right)^* \delta_{i3} \delta_{j3}
%-2 \paver{\delta w'' v^{\prime \prime *}  }   \delta \left( \frac{d  \coh{w}}{d y} \right) \delta_{i3} \delta_{j3}}_{\cs{\pph}_{33}}  + \\
+ \underbrace{ \left[ - \paver{\delta u_i''  \delta v''}        \left( \frac{\partial \coh{u}}{\partial y} \right)^* \delta_{j1} - \paver{\delta u_j''  \delta v''}        \left( \frac{\partial \coh{u}}{\partial y} \right)^* \delta_{i1} - \paver{\delta u_i''  v^{ \prime \prime *} }    \delta \left( \frac{\partial \coh{u}}{\partial y} \right) \delta_{j1} - \paver{\delta u_j''  v^{ \prime \prime *} }    \delta \left( \frac{\partial \coh{u}}{\partial y} \right) \delta_{i1}\right]}_{\cs{\pph}_{ij}} + \\
+ \underbrace{ \left[ - \paver{\delta u_i''  \delta v''}        \left( \frac{\partial \coh{w}}{\partial y} \right)^* \delta_{j3} - \paver{\delta u_j''  \delta v''}        \left( \frac{\partial \coh{w}}{\partial y} \right)^* \delta_{i3} - \paver{\delta u_i''  v^{ \prime \prime *} }    \delta \left( \frac{\partial \coh{w}}{\partial y} \right) \delta_{j3} - \paver{\delta u_j''  v^{ \prime \prime *} }    \delta \left( \frac{\partial \coh{w}}{\partial y} \right) \delta_{i3} \right]}_{\cs{\pph}_{ij}} + \\
\underbrace{+ \frac{1}{\rho} \paver{\delta p''  \left( \frac{\partial \delta u_i''}{\partial X_j} \right) }
            + \frac{1}{\rho} \paver{\delta p''  \left( \frac{\partial \delta u_j''}{\partial X_i} \right) }}_{\piph^s_{ij}}
\underbrace{-4 \epsilon_{ij}^{s*} }_{\dph^s_{ij}}.
\end{gathered}
\end{equation}

Here the mean transport term contributes to $\phiph_{x}^s$, consistently with a non-zero streamwise mean velocity $U$. Similarly, coherent transport appears in $\phiph_{x}^s$ and $\phiph_{z}^s$, since $\coh{u} \ne 0$ and $\coh{w} \ne 0$. 
Since no external volume forcing acts on the flow, the interaction forcing term is zero for both components.

\section{Analysis of conditionally-averaged quantities}
\label{sec:ens-aver}

In this Appendix, the interpretations of the local maxima of $\paver{\delta w'' \delta w''}$ in the $r_x=r_y=0$ and $r_z=r_y=0$ planes provided in \S\ref{sec:ph-by-ph} are supported by inspecting the velocity field induced by the conditionally-averaged quasi-streamwise vortex at different phases of the control cycle. 
The procedure to extract the conditional average from the DNS database closely resembles that presented by \cite{jeong-etal-1997}; it is described in detail by \cite{gallorini-quadrio-gatti-2022} and is not repeated here.

\begin{figure}
\centering
\includegraphics[width=1.0\textwidth]{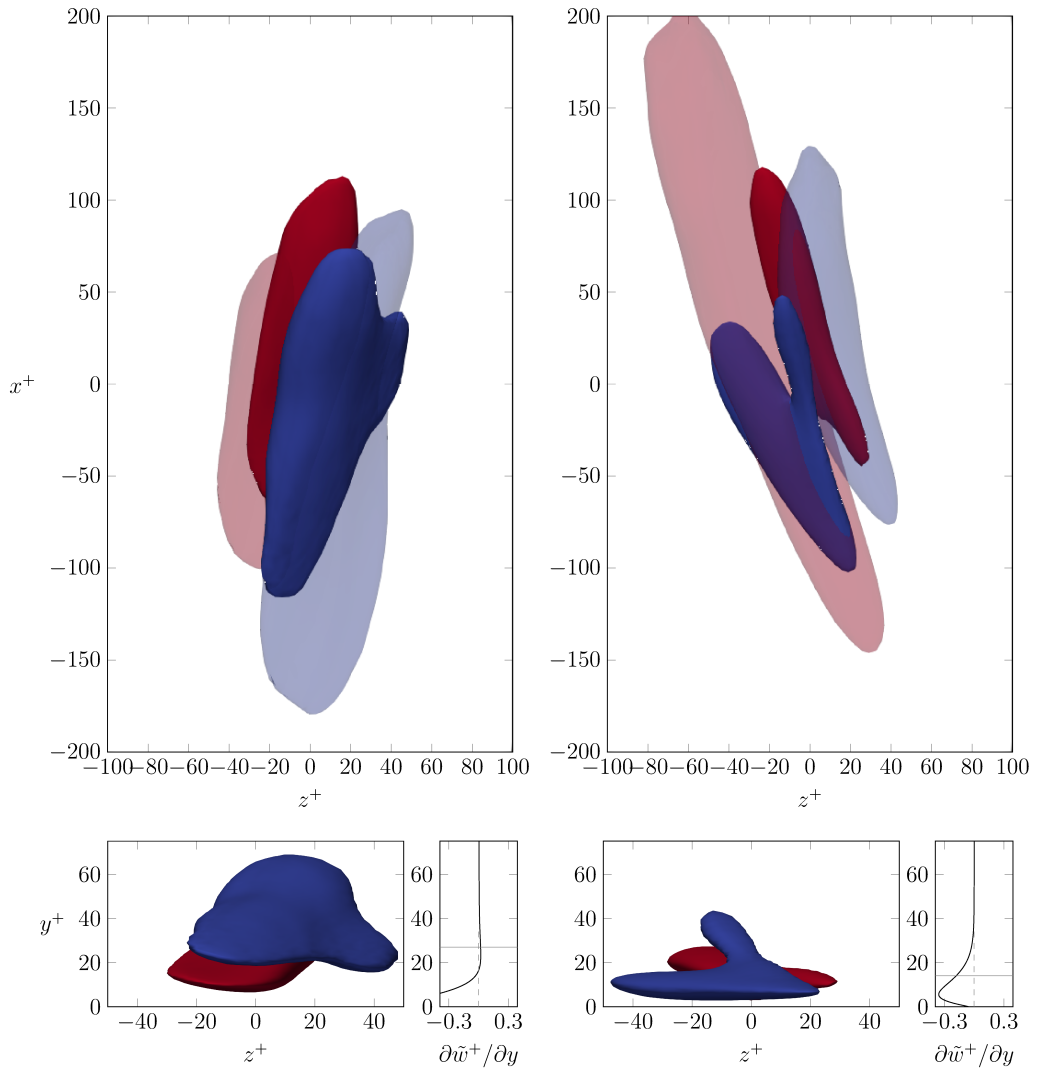}
\caption{Conditionally-averaged structure, extracted at $\varphi_1$ (left) and $\varphi_3$ (right) at $T^+=250$. The spatial shape of the structure is shown via isosurfaces of $u'^+$ (transparent color) and $w'^+$  (solid color) velocity fluctuations at the level $\pm 0.5$ (red/blue is positive/negative). The bottom panels also include the spanwise shear d$\coh{w}^+$/d$y$ at that phase, and show the wall-normal position where the extraction procedure is carried out.}
\label{fig:isosurfaces}
\end{figure}

Figure \ref{fig:isosurfaces} uses velocity isosurfaces to describe the spatial shape of the conditionally-averaged negative rotating (SN) structure for the case at $T^+=250$ at the two phases $\varphi_1$ and $\varphi_3$. 
The extraction procedure is centered at the wall-normal position of the maxima of $\paver{\delta w'' \delta w''}$ for $\varphi_1$ and $\varphi_3$ (see figure \ref{fig:energy}): this position is shown in the shear panel at the bottom of figure \ref{fig:isosurfaces}. 
At the two chosen phases, the structures show their maximum negative and positive tilt angle; however, the discussion below for $\varphi_1$ can be extended to $\varphi_2$, and that for $\varphi_3$ extends to $\varphi_4$. 
Isocontours of streamwise (transparent) and spanwise (solid color) velocities are shown in a view from above (top) and from upstream (bottom). 

Following the discussion in \S\ref{sec:ph-by-ph}, when the tilting angle is negative (see $\varphi_1$), the low-speed streak associated with a SN redistributes its energy via pressure strain and creates negative spanwise velocity fluctuations; the opposite occurs for the high-speed streak. 
This is confirmed by the ensemble-averaged structure, which shows a region of positive (negative) spanwise velocity close to the side of the high- (low-) speed streak. 
At $\varphi_3$, instead, the tilt angle of the streak is positive,
and the low- (high-) speed streak induces positive (negative) $w''$ velocity fluctuations at its side. 

Another view of the spanwise velocity contours is displayed in the bottom panels of figure \ref{fig:isosurfaces}. 
In these images, the streamwise velocity contours are removed, to focus on the spanwise component only. In the canonical channel flow, a negatively rotating vortex induces two regions of high and low spanwise velocity below and above its center, respectively. 
However, when the wall oscillates, two additional regions of positive and negative spanwise velocity originate at the sides of the tilted vortex because of its interaction with the Stokes layer.
At phase $\varphi_1$ (left panel), the peak of $\paver{\delta w'' \delta w''}$ occurs at $Y^+=25$, where the spanwise shear $\partial \coh{w}^+ / \partial y$ is positive. 
Therefore, the negatively rotating quasi-streamwise vortex lifts low spanwise velocity fluid, and displaces high spanwise velocity fluid downwards. 
This process explains the appearance of a low $w$-velocity region at the right side of the quasi-streamwise vortex, whereas the high spanwise velocity region is absorbed into the lower-side one. 
At $\varphi_3$ the regions of low/high spanwise velocity are opposite compared to $\varphi_1$ owing to the opposite sign of the spanwise shear at the location of the peak of $\paver{\delta w'' \delta w''}$ at this phase.

\section*{Declaration of Interests} 
The authors report no conflict of interest.

\section*{Author ORCIDs}
Federica Gattere, https://orcid.org/0000-0002-1450-415X\\
Alessandro Chiarini, https://orcid.org/0000-0001-7746-2850\\
Emanuele Gallorini, https://orcid.org/0000-0002-8547-3100\\
Maurizio Quadrio, https://orcid.org/0000-0002-7662-3576

\bibliographystyle{jfm}

\end{document}